\documentclass[authoryear,5p,twocolumn]{elsarticle}

\newcommand{\largetabletype}{table*}

\usepackage{graphicx}
\usepackage{amsmath, amsthm, amssymb}
\usepackage{url}
\usepackage{rotating}

\begin{document}

\title{Matching asteroid population characteristics with a model constructed from the YORP-induced rotational fission hypothesis}
\author[label1,label1b,label1c]{Seth A. Jacobson}
\author[label2]{Francesco Marzari}
\author[label3]{Alessandro Rossi}
\author[label4]{Daniel J. Scheeres}
\address[label1]{Department of Astrophysical and Planetary Sciences, University of Colorado, Boulder, CO 80309-0391, USA}
\address[label1b]{Laboratoire Lagrange, Observatoire de la C{\^o}te d'Azur, Boulevard de l'Observatoire, 06304 Nice Cedex 4, France}
\address[label1c]{Bayerisches Geoinstitut, Universtat Bayreuth, D-95444 Bayreuth, Germany}
\address[label2]{Dipartimento di Fisica, Universit{\`a} di Padova, 35131 Padova, Italy }
\address[label3]{IFAC-CNR, 50019 Sesto Fiorentino, Firenze, Italy}
\address[label4]{Department of Aerospace and Engineering Sciences, University of Colorado, Boulder, CO 80309-0429 USA}

\begin{abstract}
From the results of a comprehensive asteroid population evolution model, we conclude that the YORP-induced rotational fission hypothesis can be consistent with the observed population statistics of small asteroids in the main belt including binaries and contact binaries.
The foundation of this model is the asteroid rotation model of \citet{Marzari:2011dx}, which incorporates both the YORP effect and collisional evolution.
This work adds to that model the rotational fission hypothesis, described in detail within, and the binary evolution model of \citet{Jacobson:2011eq,Jacobson:2011hp}.
The asteroid population evolution model is highly constrained by these and other previous works, and therefore it has only two significant free parameters: the ratio of low to high mass ratio binaries formed after rotational fission events and the mean strength of the binary YORP (BYORP) effect.

We successfully reproduce characteristic statistics of the small asteroid population: the binary fraction, the fast binary fraction, steady-state mass ratio fraction and the contact binary fraction.
We find that in order for the model to best match observations, rotational fission produces high mass ratio ($> 0.2$) binary components with four to eight times the frequency as low mass ratio ($< 0.2$) components, where the mass ratio is the mass of the secondary component divided by the mass of the primary component.
This is consistent with post-rotational fission binary system mass ratio being drawn from either a flat or a positive and shallow distribution, since the high mass ratio bin is four times the size of the low mass ratio bin; this is in contrast to the observed steady-state binary mass ratio, which has a negative and steep distribution. 
This can be understood in the context of the BYORP-tidal equilibrium hypothesis, which predicts that low mass ratio binaries survive for a significantly longer period of time than high mass ratio systems.
We also find that the mean of the log-normal BYORP coefficient distribution $\mu_B \gtrsim 10^{-2}$, which is consistent with estimates from shape modeling \citep{McMahon:2012ti}.
\end{abstract}

\maketitle

\section{Introduction}
\label{sec:introduction}
The YORP-induced rotational fission hypothesis predicts that the Yarkovsky-O'Keefe-Radzievksii-Paddack (YORP) effect can rotationally accelerate rubble pile asteroids until internal stresses within the body due to centrifugal accelerations surpass the gravitational attractions holding the rubble pile elements in their current configurations.
Subsequently, according to the hypothesis, these asteroids rotationally fission into mutually orbiting components that can dynamically evolve into the observed binary populations \citep{Bottke:2006en,Scheeres:2007io,Walsh:2008gk,Jacobson:2011eq}.
This hypothesis has been constructed on two pillars: the theoretical conclusion that light imparts a meaningful torque on small asteroids, which has been named the YORP effect \citep{Rubincam:2000fg}, and the observations that the majority of binary asteroid systems have rapidly rotating primaries and small semi-major axes relative to the radius of the primary.
This configuration has a high angular momentum content, which is consistent only with formation from rotational fission \citep{Margot:2002fe}.

The hypothesis has also undergone observational and theoretical experiments.
Rotational fission predicts a relationship between the angular momentum content of the fissioned asteroid system and the mass ratio between its components \citep{Scheeres:2007io}.
In the asteroid pair population, \citet{Pravec:2010kt} discovered that the spin rates of the larger members and the mass ratio of each observed asteroid pair had the predicted relationship.
This confirmed that asteroid pairs are the result of rotational fission.
\citet{Jacobson:2011eq} tested the connection between rotational fission and the observed binary population by numerically modeling the post-rotational fission process.
With only the inclusion of gravitational dynamics and mutual body tides, they were able to create the most commonly observed asteroid systems (e.g. asteroid pairs, binaries, contact binaries, etc.).
After including the binary YORP (BYORP) effect, all the observed binary systems are hypothesized to be natural final states after these processes (as reviewed in \citet{Jacobson:2014hp}).

Often asteroid evolution occurs too quickly (on Solar System timescales) and too infrequently (on human timescales) to be observed {\it in situ}.
Although, as larger telescopes are aimed at smaller asteroid systems, the possibility of capturing rotational fission events as they occur grows increasingly high \citep{Marzari:2011dx} (and the first such systems may have already been observed, e.g. \citet{Jewitt:2010bs} and \citet{Jewitt:2014fe}).
In the meantime, these timescales present a challenge for direct confirmation of rotational fission and subsequent binary evolution, but the proposed asteroid evolution makes specific predictions for the relative abundances of each final state so a detailed asteroid population evolution model that reproduces the observed sub-populations is a strong consistency test of the YORP-induced rotational fission hypothesis.
We present such an asteroid population evolution model that allows us to see if the proposed evolutionary mechanisms are sufficient to create the observed sub-populations and, perhaps more importantly, create them in the right proportions to one another.

The asteroid population evolution model is a development of the model presented in \citet{Marzari:2011dx}, which studied the rotational evolution of the Main Belt asteroid (MBA) population including both the YORP effect and collisions.
This model was already an improvement and continuation of earlier projects by \citet{Rossi:2009kz} and \citet{Scheeres:2004bd}, which studied the near-Earth asteroid population.
Similar to \citet{Marzari:2011dx}, we use a Monte Carlo approach to simulate the evolution of $2 \times 10^6$ asteroid systems for $4.5 \times 10^9$ years.
The spin state of each asteroid evolves constantly due to the YORP effect and collisions as in \citet{Marzari:2011dx} (summarized in Section~\ref{sec:singleasteroidevolution}).
Similar to \citet{Jacobson:2014bi}, when the rotation rate of an asteroid exceeds a specified spin limit, the asteroid rotationally fissions and can form a binary system.
The survival and lifetimes of these binary systems are determined from a separate set of calculations based on the results of \citet[][b]{Jacobson:2011eq}.

Both the single and binary evolution schemes are built from well-developed theories in the literature.
Therefore, there are very few free parameters built into the model that have not been significantly constrained elsewhere.
For instance, the intrinsic probability of collision for Main Belt asteroids $\left< P_i \right> = 2.7 \times 10^{-18}$ yr$^{-1}$ km$^{-2}$, the fundamental parameter determining the frequency of collisions in the model, has been established by the efforts of a series of authors to at least the order of uncertainty inherent in other parts of the asteroid population evolution model \citep{Farinella:1992im,BottkeJr:1994kr}.
Similarly, the binary evolution model utilizes the the evolutionary flowchart and derived probabilities given in \citet[][b]{Jacobson:2011eq}.

The binary evolution model does contain two free input parameters that are not well constrained by either observation or current theory.
The first parameter is the initial mass ratio fraction $F_i$, which is the ratio of high mass ratio to low mass ratio binary systems created from rotational fission events.
This parameter is determined from the interior structure of the rotationally fissioning asteroid and the mechanics of the fission event itself, neither of which are currently observed or modeled accurately enough to generate this number.
The initial mass ratio fraction is distinct from the observed mass ratio fraction $F_q$, which reflects the evolutionary differences between high and low mass ratio systems.

The second parameter is the mean of the logarithmic normal distribution of the BYORP coefficient $\mu_B$.
It is used to determine the strength of the BYORP effect, which determines the bound lifetimes for most binary systems.
The basic shape and width of the distribution is determined from the equilibrium occupied by the synchronous binary asteroid population.
There has only been a single published estimation of a BYORP coefficient and the shape model used may not have had the necessary accuracy \citep{McMahon:2010jy} and this effect has yet to be measured.
These parameters are the knobs that will control the output from the asteroid population evolution model.

After evolving the population for the age of the Solar System, which is longer than needed for the sub-populations to reach a relative steady-state equilibrium for most choices of $\mu_B$, we can compare the model to the observed main asteroid belt.
There are four particular observables that we can compare with our model: The binary fraction $F_B$, which is the total number of binaries over the total number of asteroid systems, the fast-rotating binary fraction, $F_F$, which is a more specific comparison of the number of binaries with rapidly rotating primaries to the number of rapidly rotating asteroids, the steady-state (i.e.
observed) mass ratio fraction $F_q$, which is defined similarly to the initial mass ratio fraction $F_i$ above, and the contact binary fraction $F_C$, which is the number of contact binaries over the total number of asteroid systems.
From these comparisons, we construct a simple log-likelihood model to assess which model parameters, $F_i$ and $\mu_B$, are the most likely to match the model population to the observations.
Lastly, we discuss the best fit models and their implications for future observations and tests.

\section{Single Asteroid Evolution}
\label{sec:singleasteroidevolution}
Each asteroid within the asteroid population evolution model is individually evolved.
Similar to \citet{Marzari:2011dx}, the asteroid population evolution model utilizes the intrinsic probability for impact $\left< P_i \right>$ and a projectile size frequency distribution to determine the collision history of each model asteroid.
Between collisions, single asteroids undergo rotational evolution driven by the YORP effect, which modifies both the spin rate and obliquity of the asteroid.
Rotational acceleration can lead to rotational fission if it occurs before the next collision event.
The specific conditions for triggering rotational fission and the process itself are parameterized using well-developed models \citep{Scheeres:2007io,Jacobson:2011eq}.

Each asteroid system is characterized by a number of fixed and evolving parameters.
These parameters change if the system undergoes rotational fission and evolves into a binary asteroid system.
All systems are assigned a fixed semi-major axis $a_\odot$ and eccentricity $e_\odot$ from a Main Belt asteroid orbital element distribution.
Both the YORP effect and collisions evolve the spin rate $\omega$ and the obliquity $\epsilon$ of each asteroid.
The initial spin rate is drawn from a Maxwellian distribution with a $\sigma = 1.99$ corresponding to a mean period of $7.56$ hrs, which is consistent with \citet{Fulchignoni:1995um} and \citet{Donnison:1999iv}. \citet{Rossi:2009kz} demonstrated for models similar to the asteroid population evolution model that the steady-state spin rate distribution is independent of the initial spin rate distribution.
We draw the initial obliquity of each asteroid from a flat distribution.
The relative change in obliquity is used by the model to update the YORP coefficient, however the absolute obliquity is not currently used by the model.
Thus the rotational evolution output is insensitive to the initial obliquity distribution, but it is a feature of the model that could be utilized in the future to compare input and output obliquity distributions.

For the purpose of calculating the critical spin limit, each asteroid is assigned a shape from an ellipsoidal semi-axis ratio distribution reported from laboratory experiments by \citet{Giblin:1998io}.
From largest to smallest the tri-axial semi-axes are $a$, $b$, and $c$.
Axis ratios are drawn from normal distributions such that for $b/a$, the mean $\mu = 0.6$ with a standard deviation $\sigma = 0.18$ and for $c/a$, $\mu = 0.4$ and $\sigma = 0.05$.
This shape distribution is in agreement with Hayabusa observations of boulders on 243 Itokawa and photometry of small, fast-rotating asteroids \citep{Michikami:2010cr} and the mean lightcurve amplitude of small asteroids with diameters between 0.2 and 10 km \citep{Pravec:2000dr}.
The added realism of using a shape distribution rather than assuming sphericity results in a reduced critical spin limit.

The most important parameter for determining the evolution of a individual asteroid system is its mean radius $R$.
Both the collision and rotational evolution depend strongly on the size of the asteroid, thus the two effects are not of comparable strengths at all radii.
We expect radiative torques to be inconsequential for large asteroids, but dominant at smaller sizes transitioning at some critical radius $R_c$.
This critical radius is estimated to be $R_c \approx 6$ km from both analytical arguments and numerical experiments \citep{Jacobson:2014bi}.
In Section~\ref{sec:rotationalspinlimits}, we discuss a transition between ``monolithic'' and ``rubble pile'' interior structures that is inferred to occur at a radius of $R \approx 125$ m.
Therefore the asteroid population evolution model focuses on asteroids with radii between $R = 200$ m, just above this transition, and $20$ km, since the evolution of asteroids with radii $R \gtrsim 20$ km are collision dominated and have YORP effect timescales on order with the age of the Solar System or longer \citep{Jacobson:2014bi}.
Within this range, the asteroid population evolution model includes a sample of 2 million asteroids which are drawn from the size frequency distribution derived from the results of the Sloan Digital Sky Survey \citep{Ivezic:2001ct}.
The range of asteroids included in the asteroid population evolution model is different than the range of possible projectile asteroids used to model collisions.
These projectiles range in radius from $0.05$ m to $20$ km.

Asteroid system destruction whether through a catastrophic collision, rotational bursting, or destruction of a binary, is a mass transfer from one size asteroid (the progenitor in the case of a binary) into two or more smaller size bodies.
Each asteroid in the asteroid population evolution model resides in a logarithmic diameter bin and the model tracks this mass flow from larger bins into smaller bins after each destructive event.
Diameter bins are created so that the upper diameter of a bin is $D_i = D_m D_w ^i$, where $D_m$ is the minimum diameter and $D_w = 1.25992$ is the bin width.
After a destructive event the asteroid within the asteroid population evolution model is replaced with another asteroid from the original diameter bin.
This replacement is motivated by the constant flux of material into the original bin from even larger bins, and in this way, the asteroid population evolution model maintains a steady-state size frequency distribution.
Therefore it does not feature a full feedback size-frequency distribution.
The output of the asteroid population evolution model includes destruction statistics that we published in \citet{Jacobson:2014bi} to generate a new size frequency distribution.
The asteroid population evolution model is utilized here to test the YORP-induced rotational fission hypothesis, so rather than focus on changes to the size-frequency distribution, we focus on the abundances of distinguishable sub-populations taking into account both collisional and rotational evolution.

\subsection{Collisional Evolution}
\label{sec:collisionalevolution}
The collisional evolution of each asteroid follows a similar protocol as \citet{Marzari:2011dx}.
The population of potential impactors is derived from the Sloan Digital Sky Survey size frequency distribution of asteroids \citep{Ivezic:2001ct} distributed over logarithmic size bins from $0.05$ m to $20$ km.
Using Poisson statistics, the number of collisions and their timing is computed for each asteroid with projectiles from each size bin using the intrinsic probability of collision for the Main Belt $\left< P_i \right> = 2.7 \times 10^{-18}$ km$^{-2}$ yr$^{-1}$ \citep{Farinella:1992im}.
Each collision is assigned an impact velocity of $5.5$ km s$^{-1}$ \citep{BottkeJr:1994kr} and a random geometry within the limits of the Main Belt orbital distribution, in order to determine from these parameters the change in spin rate due to each collision.

Using this method, we have created a list of collisions and their properties for each asteroid in our simulated population.
Between each collision in the list, each asteroid rotationally evolves according to the YORP effect.
At the time of a collision the rotational evolution is stopped and the collision is evaluated.
First, the collision is classified as either a cratering or a catastrophic collision depending on the energy of the event.

If the collision is too large for a cratering event, then the original asteroid is shattered and a new object is created with the same size but a new initial spin state and YORP coefficient.
Shattering collisions are defined as those that deliver specific kinetic energy greater than the critical specific energy of the target, which defined as the energy per unit target mass delivered by the collision required for catastrophic disruption (i.e.
such that one-half the mass of the target body escapes).

Cratering collisions do not appreciably change the mass or size of the target asteroid, but they can change the angular momentum of the asteroid.
The angular momentum of the projectile, the target and the geometry of the collision determine the new angular momentum of the cratered asteroid.
This new angular momentum vector is used to update both the spin rate and the obliquity.
The model neglects the angular momentum removed by fragments.
This assumption is acceptable for the frequent low energy impacts but introduces a small error for high energy impacts that do not catastrophically disrupt the asteroid, which are infrequent.
Sub-catastrophic impacts create a random walk in spin rate if there is no significant YORP effect rotational acceleration.

\subsection{YORP Evolution}
\label{sec:yorpevolution}
The YORP effect changes the spin rate $\dot{\omega}$ as \citep{Scheeres:2007kv}:
\begin{equation}
\dot{\omega} = \frac{Y}{2 \pi \rho R^2} \left( \frac{F_\odot }{a_\odot^2 \sqrt{1 - e_\odot^2}} \right)
\end{equation}
where $F_\odot = 10^{14}$ kg km s$^{-2}$ is the solar radiation constant and $Y$ is a non-dimensional YORP coefficient $Y$ assigned to each object from a gaussian distribution with a mean of $0$ and a standard deviation of $0.0125$ which was found to successfully reproduce the spin rate distribution of both the near-Earth and main belt asteroid populations \citep{Rossi:2009kz,Marzari:2011dx}.
In \citet{Rossi:2009kz}, the results were found to be invariant on the order of the uncertainty of the model to the particular distribution used.
This distribution is consistent with the measured values of 1862 Apollo (1932 HA) $Y = 0.022$ \citep{Kaasalainen:2007hq} and 54509 YORP (2005 PH$_5$) $Y = 0.005$ \citep{Taylor:2007kp,Lowry:2007by}.
The model does not distinguish the Tangential YORP effect \citep{Golubov:2012kt,Golubov:2014hf}, which may both bias the sense of rotation (towards prograde) and increase the acceleration when the asteroid is rotating slowly.
The first effect cannot be captured in the model since it does not track sense of rotation but the second effect is already empirically included in the model, since the utilized YORP coefficient distribution was successfully able to reproduce the asteroid spin rate distribution.

Using semi-major axis drift as a proxy for obliquity evolution, \citet{Bottke:2015jg} hypothesized that obliquity must be preserved through multiple rotational fission events or the rotational fission timescale must be suppressed, which they accomplished via a stochastic YORP.
This model effectively includes the effect of stochastic YORP on rotation rate evolution since the YORP coefficient is re-drawn after each rotational fission event, significant collisions and whenever the obliquity changes by more than $0.2$ rad due to either collisions or YORP evolution itself.
For smaller changes in the obliquity, the YORP coefficient evolves according to: $Y' =  Y \left( 3 \cos^2 \epsilon - 1 \right) / 2$ as in \citet{Nesvorny:2008by}.
The model only tracks this relative obliquity evolution due to the YORP effect and is not a full obliquity evolution model.
How the obliquity evolves after a rotational fission occurs is not clear, and the role of binary formation and evolution on obliquity has not been fully explored.

If the YORP coefficient $Y > 0$, then the spin rate is accelerating and if uninterrupted by collisions, the spin rate will eventually reach the spin limit.
If the YORP coefficient $Y < 0$, then the spin rate is decelerating.
Eventually, if uninterrupted by collisions, the angular momentum of the asteroid becomes very low, where even the smallest projectiles can deliver impulsive torques that are the same order of magnitude as the angular momentum of the target body.
Since this model cannot assess the evolution of this state, i.e. we only model the effects of $0.05$ m projectiles and larger, an artificial lower spin barrier at 10$^5$ hr is enforced and at this very slow rotation rate the YORP torque switches directions.
This assumption could underestimate the YORP evolution timescale but by less than a couple thousand years for even the largest asteroids in the model.

\subsection{Spin Limits}
\label{sec:rotationalspinlimits}
Almost all asteroids larger than approximately 200 m in diameter obey a critical disruption spin limit of about 2.3 hours \citep{Pravec:2000dr}.
Below this size, there are a couple hypotheses for why the barrier can be broken including enhanced strength due to cohesive forces \citep{Scheeres:2012tj, Holsapple:2007eg} and that these super-critical asteroids are the monolithic remnants of rubble pile progenitors that have undergone multiple YORP induced rotational fissioning \citep{Pravec:2007ki}.
To limit the complexity of the model, we only consider asteroids with radii $R \geq 0.2$ km.

The critical disruption spin limit is a direct consequence of the YORP-induced rotational fission hypothesis.
As an asteroid is rotationally accelerated due to either a continuous YORP torque or sudden collisional torque the centrifugal accelerations increase on each component of a  rubble pile asteroid.
These accelerations counter the gravitational accelerations holding the body together.
\citet{Scheeres:2009dx} showed that for every partitioning of the body in two along rubble pile component boundaries, there is a specific rotation rate at which the centrifugal accelerations will exceed the mutual gravity and the two sections will no longer rest against each other but enter into orbit.
As the body rotationally accelerates it will reach the slowest of these rotation rates first and it will be along this partitioning that the body rotationally fissions.
The smaller of the two sections is now the secondary, and the remainder is the primary, both in orbit about each other.
This simple story of rotational fission is complicated by but reaffirmed when the asteroid's shape is also allowed to evolve.
Some numerical models predicts surface shedding implying a very low initial mass ratio fraction \citep{Walsh:2008gk,Hirabayashi:2015jd}, while others predict internal failure consistent with a high initial mass ratio fraction \citep{Sanchez:2012hz}.
Because of this uncertainty, the initial mass ratio fraction is a fundamental parameter of the asteroid population evolution model.
Once the initial mass ratio of a particular asteroid has been chosen, the model utilizes the simple approximation that all rubble piles rotationally disrupt at the critical disruption spin limit modified to account for the ellipsoidal shape of the asteroid.

We also consider collision-induced rotational fission, which requires that the combined angular momentum from both precursor bodies and the cratering impact geometry exceeds the critical angular momentum necessary for the body to gravitationally hold itself together against centrifugal accelerations.
This is similar to the YORP-induced rotational fission hypothesis described above with three exceptions.
Firstly, the collision may significantly change the internal component distribution itself.
Secondly, the torque is delivered impulsively.
These first two differences are not significant since we are not modeling the internal component distribution nor are we resolving the rotational fission event itself.
Thirdly, the new system angular momentum may exceed the critical angular momentum by a measurable amount.
Even though an asteroid that undergoes collision-induced rotational fission may be rotationally accelerated past the critical disruption rotation rate, for the purposes of the asteroid population evolution model these events will be treated the same  as the YORP-induced rotational fission, which occurs at the critical disruption rotation rate.
Consequences of ignoring the excess include overestimating the binary creation rate at the expense of the asteroid pair creation rate.

\subsection{Outcomes of Rotational Fission}
\label{sec:outcomesofrotationalfission}
\begin{figure}[!tb]
\centering
\includegraphics[width=\columnwidth]{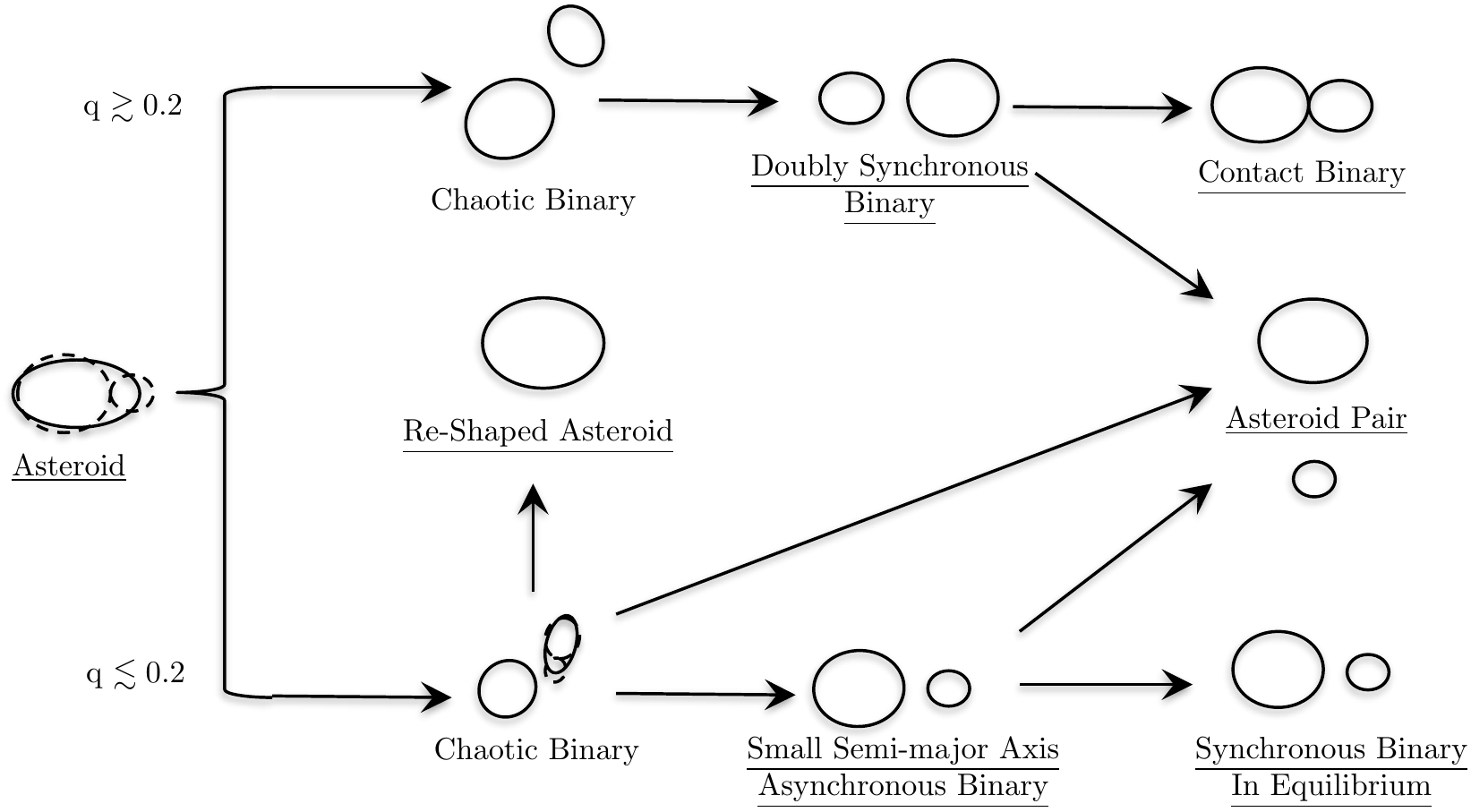}
\caption{Evolutionary tracks for a small asteroid after it has undergone rotational fission according to the theory in \citet{Jacobson:2011eq} and \citet{Jacobson:2011hp}.
Each evolutionary step is indicated by an arrow.
Most of this diagram is a cycle, since the end states are single asteroids: re-shaped asteroids, contact binaries or each member of asteroid pairs.
Collisions can destroy synchronous binaries in equilibrium.}
\label{fig:AsteroidFlowChart}
\end{figure}
If the critical spin rate is reached, then the asteroid population evolution model simulates a rotational fission event for that asteroid.
This can happen when a collision brings the asteroid above the rotational breakup limit or when the rotational breakup period is reached due to YORP acceleration.
\citet{Pravec:2010kt} observationally showed that these types of events are the progenitors of the observed asteroid pair population.
\citet{Jacobson:2011eq} numerically showed that rotationally fissioned asteroid systems can evolve into a number of different outcomes, as shown in Figure~\ref{fig:AsteroidFlowChart}, but the chaotic nature of the system allows for only a probabilistic determination of the outcome.
A binary system formed via rotational fission can temporarily occupy a number of evolutionary morphologies before settling into three enduring states: single, binary and pair.
None of these categories are truly permanent since single asteroids can undergo rotational fission forming binaries and pairs, binaries can be disrupted forming pairs or collide to make re-shaped asteroids (i.e. singles), and asteroid pairs, which are really single asteroids, can be rotationally fissioned.

The mass ratio, which is the mass of the primary divided by the mass of the secondary, determines the energy available to the post-fission binary system \citep{Scheeres:2009dc}.
If the mass ratio $q > 0.2$, then the system has a negative free energy and so is bound.
These binaries cannot form asteroid pairs without an external force or torque such as the YORP effect.
Whereas, systems with mass ratios $q < 0.2$ are unbound systems with positive energy, and so can immediately disrupt to form asteroid pairs \citep{Pravec:2010kt}.
Because of this fundamental difference, high mass ratio $q > 0.2$ and low mass ratio $q < 0.2$ binary systems evolve differently within the model.

\citet{Jacobson:2011eq} determined that the mass ratio is not necessarily a fixed quantity and may change via a process termed secondary fission.
During secondary fission, the secondary undergoes rotational fission similar to that which formed the binary system in the first place with the exception that the rotational torque is provided by spin-orbit coupling rather than the YORP effect .
This process was only observed numerically to occur with low mass ratio systems and since it reduces the mass ratio, no binary systems can evolve across the $q \sim 0.2$ threshold between high and low mass ratio systems.

\subsection{Mass Ratio Fraction}
\label{sec:massratiofraction}
Before describing the possible outcomes and their likelihoods for both high and low mass ratio systems, the relative number of high to low mass ratio systems must be determined.
The initial mass ratio of a binary system after rotational fission is determined by the internal component (i.e. rubble pile element) distribution of the parent asteroid before rotational fission \citep{Scheeres:2007io}, so it is the distribution of internal structures amongst an ensemble of asteroids that will determine the initial distribution of binary mass ratios.
The direct determination of the distribution of mass ratios after rotational fission would perhaps require the gentle and complete disassembly of a number of asteroids into their component pieces understanding their masses, shapes and relative locations, however an approximate understanding of this distribution may soon be available via detailed numerical modeling using discrete element methods \citep{Walsh:2008gk,Walsh:2012jt,Sanchez:2011kw,Sanchez:2012hz}.

\begin{figure}[!tb]
\centering
\includegraphics[width=\columnwidth]{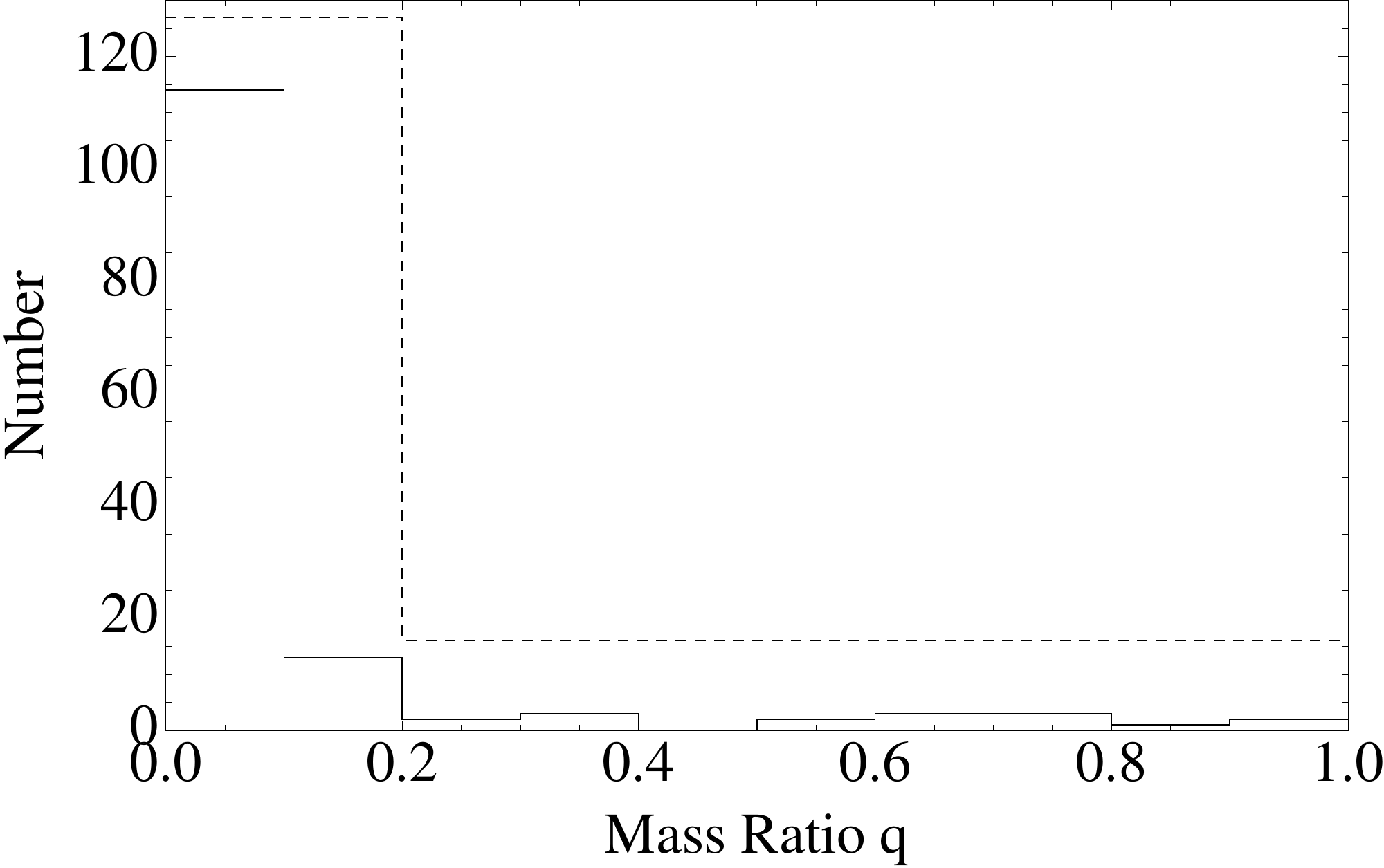}
\caption{Two histograms of the same observed binary distribution as a function of mass ratio.
The solid histogram shows the number of binaries in bins of width $0.1$ in mass ratio.
The dashed histogram simply outlines the number of binaries in the low mass ratio ($0 < q < 0.2$) and the high mass ratio  ($0.2 < q < 1$) populations, of which there are $127$ and $16$ observed binary systems, respectively.
The observed binaries are the 143 characterized binaries with small primary diameters $\lesssim15$ km according to the September 18, 2015 binary asteroid parameter release from http://www.asu.cas.cz/~asteroid/binastdata.htm as compiled by methods and assumptions described in \citet{Pravec:2006bc} and updated in \citet{Pravec:2015uc}.}
\label{fig:BinaryMassRatioHist}
\end{figure}

Until then, we can constrain the initial mass ratio fraction $F_i$ that is input in the asteroid population evolution model by comparing the observed steady-state mass ratio fraction to the steady-state fraction output by the model $F_q$.
The steady-state distribution reflects a balance between creation and destruction of binary systems as a function of mass ratio.
The mass ratio fraction $F$ is defined as the number of high mass ratio systems divided by the number of low mass ratio systems.
The mass ratio fraction is a function of time as high and low mass ratio systems are created and destroyed.
The initial mass ratio fraction $F_i$ reflects the distribution of possible internal component distributions of parent asteroids.
This initial distribution then evolves into the observed steady-state mass ratio fraction $F_q$ due to the differences between binary creation and destruction timescales in high and low mass ratio systems.
The initial mass ratio fraction $F_i$ is an input into the asteroid population evolution model, and the steady-state mass ratio fraction $F_q$ is one of the observable outputs.

This evolution in mass ratio fraction is due only to the creation and destruction of specific binary systems and not due to the possible evolution in mass ratio of those systems, since high mass ratio systems were not observed in numerical models to transform into low mass ratio binaries and vice versa \citep{Jacobson:2011eq}.
As discussed above, binary systems cannot cross the mass ratio $q \sim 0.2$ boundary between the two regimes via secondary fission.

Thus, the simplest approximation within each mass ratio regime is to assume that the members are selected from a flat distribution.
As is shown in Figure~\ref{fig:BinaryMassRatioHist}, this description is imperfect but is an appropriate assumption, since the asteroid population evolution model is only being used to determine the steady-state mass ratio fraction $F_q$ and not the detailed steady-state mass ratio distribution.
In the future, a treatment that includes a more advanced binary evolution model with a more detailed dependance on mass ratio will also need to explore more complex initial mass ratio distributions.

The range of initial mass ratio fractions $F_i$ to be tested in the asteroid population evolution model is motivated by the observed population as shown in Figure~\ref{fig:BinaryMassRatioHist}.
The observed steady-state mass ratio fraction is $F_q \sim 0.2$, but low mass ratio systems face much steeper odds of surviving as binary systems ($8\%$), as discussed in Section~\ref{sec:instantaneousbinaryevolution}.
To examine a broad range of initial conditions and their outcomes: $F_i$ is varied between $32$, $16$, $8$, $4$, $2$, and $1$.
Every time a binary system is created via rotational fission in the asteroid population evolution model, the binary is assigned to either the low or high mass ratio regime, such that $\left( 1 + F_i \right)^{-1}$ of the time the system is low mass ratio and $1 - \left( 1 + F_i \right)^{-1}$ of the time it is high mass ratio.
This is the first knob in the model as described in Section~\ref{sec:introduction}; the other knob is the BYORP coefficient distribution.

\section{Binary Asteroid Evolution}
\label{sec:binaryasteroidevolution}
After a rotational fission event, a binary system is formed that undergoes complex dynamics immediately after formation \citep{Jacobson:2011eq}.
If they stabilize, then non-gravitational and tidal torques control the fate of the system \citep[e.g.][]{vanFlandern:1979tf,Cuk:2005hb}.
Since this evolution is complex, the asteroid population evolution model does not individually evolve binary systems, since this would be computationally expensive.
Instead, a lifetime for each system is drawn from a distribution, which has been determined from a separate Monte Carlo model of binary asteroid evolution as described later in this section.
After formation each binary system is placed randomly in a mass ratio bin according to the probabilities established by the initial mass ratio fraction $F_i$: low ($q < 0.2$) or high ($q > 0.2$).
These mass ratio bins determine the ``instantaneous'' survival of the binary system.
If the binary survives, then the binary's ``long-term'' evolutionary path is drawn, which is also dependent on the assigned mass ratio bin.
Each evolutionary path is associated with a binary lifetime distribution.
The drawn lifetime is then scaled by the heliocentric orbit of the system and the absolute size of the system (radius of the primary).
The heliocentric semi-major axis and eccentricity remain the same as the rotationally fissioned progenitor.

Each binary system has four permanent parameters: the heliocentric semi-major axis and eccentricity, the mass ratio and the binary lifetime.
The evolved parameter is not the spin rate as in the single asteroid case, but rather the age of the binary.
The final outcome of the evolutionary path is also recorded, so that when the binary lifetime is over, the system is replaced with a new asteroid the same size as the progenitor but labeled as either an asteroid pair or a re-shaped asteroid.
This evolution may be interrupted by a collision, and this is discussed in Section~\ref{sec:binariesandcollision}.

The evolution of a binary asteroid system from rotational fission to a long term stable outcomes is deterministic but the evolution is chaotic and only weakly a function of the shape of each body and the mass ratio within each of two distinct dynamical regimes: low and high mass ratio \citep{Jacobson:2011eq}.
The initial evolution of the spin and orbit states of the system are controlled by dynamical coupling between the spin and orbit by non-Keplerian gravity terms and solar gravitational perturbations.
This dynamical evolution is quick often finishing in tens of years \citep{Jacobson:2011eq}.
Due to the chaotic and swift nature of this evolution it occurs ``instantaneously'' and probabilistically within the model.
If the rotational fission event results in the creation of a re-shaped asteroid or an asteroid pair, then these objects are returned to the asteroid population evolution model as single asteroids sharing the same heliocentric orbit properties as their progenitors.
If the systems settles into a stable (i.e. long-lasting) binary state, then the binary evolves according to ``long-term'' binary evolution.

According to the binary evolution model described in \citet{Jacobson:2011hp}, the longevity of a binary system is primarily determined by the strength of the BYORP effect \citep{Cuk:2005hb,McMahon:2010jy}.
The BYORP effect may permanently stabilize some binaries in a tidal-BYORP equilibrium and expand the orbits of others.
Low mass ratio binaries that evolve into a tidal-BYORP equilibrium exist until a collision occurs that is capable of disrupting the mutual orbit or catastrophically destroys one of the binary members.
For other stable binary systems, after creation each is assigned a lifetime that is drawn from a distribution determined by Monte Carlo modeling of binary asteroids as explained in Section~\ref{sec:binarylifetimedistributions}.
During this evolution, binary destruction via collision is possible as discussed in Section~\ref{sec:binariesandcollision}.
At the end of a binary system's lifetime, the binary disrupts forming a re-shaped asteroid, if the BYORP effect is contractive, or an asteroid pair, if the BYORP effect is expansive.
Here, we assert that the BYORP effect can expand the mutual orbit to the Hill sphere creating an as-yet-unobserved population of asteroid pairs \citep{Jacobson:2015wu}.
However, it is possible that solar perturbations or libration growth due to the adiabatic invariant relationship between libration and mean motion de-synchronize the synchronous binary member, which is undergoing the BYORP effect \citep{Jacobson:2014hp}.
In the case that both binary members become asynchronous at a wide semi-major axis, the binary mutual orbit can no longer significantly evolve due to tides and the BYORP effect and the secondary is unlikely to be re-captured into synchronicity.
In the model, we treat this scenario identically to the formation of an asteroid pair, since the primary spin state evolves according to the YORP effect with negligible influence from the secondary because of the wide orbit.

\subsection{``Instantaneous'' Binary Evolution}
\label{sec:instantaneousbinaryevolution}
Each system that rotationally fissioned undergoes binary evolution.
Within the Monte Carlo asteroid evolution program, there are two stages for binary evolution: ``instantaneous'' and ``long-term.''
This distinction is made between processes that occur immediately after rotational fission and last less than $10^5$ years, and those that take significantly more than $10^5$ years.
This timescale was chosen since it is a tenth of the YORP timescale for an asteroid with 200 m radius at 2.5 AU, and so it is effectively the time resolution of the code.
``Instantaneous'' evolution is described below and ``long-term'' evolution in Section~\ref{sec:longtermbinaryevolution}.

An example of an instantaneous process is tidal synchronization, which has been estimated from first principles to take between 10$^3$ and 10$^5$ years for representative binaries \citep{Goldreich:2009ii}.
While the YORP effect can delay tidal synchronization \citep{Jacobson:2014jw}, tides typically dominate the spin evolution for newly created binary systems with semi-major axes less than 16 primary radii \citep[see Figure 1;]{Jacobson:2014hp}, which is the maximum distance obtained by simulated post-fission binaries \citep{Jacobson:2011eq}.
Due to spin-orbit coupling, the timescale for tidal synchronization may be lengthened since spin locking cannot occur above a specific eccentricity given the shape of the secondary \citep{Naidu:2015gp}.
Understanding the details of tidal evolution is an ongoing focus of research, for instance, if the singly synchronous binary asteroids occupy a tidal-BYORP equilibrium \citep{Jacobson:2011hp} as 1996 FG$_3$ does \citep{Scheirich:2015ez}, then tidal timescales are much shorter than those estimated purely from theory \citep{Fang:2012fw}.
Furthermore, the first-order classical constant tidal parameter ratio $k/Q$ theories are likely not correct and, for instance, they may not depend on the mechanical rigidity as assumed by many \citep{Goldreich:2009ii,Taylor:2011bj} but instead on an effective viscosity \citep{Efroimsky:2015ia} or surface properties including surface motion and potential lofting \citep{Fahnestock:2009en,Harris:2009ea}.
In the asteroid population evolution model, consistently mis-estimating the length of  ``instantaneous'' processes is effectively a bias on the determined mean of the log-normal BYORP coefficient distribution $\mu_B$.

During ``instantaneous'' evolution, the mass ratio of the newly formed binary systems is chosen randomly according to the initial mass ratio fraction $F_i$ distribution.
If the mass ratio of a system is chosen to be high, then that system evolves along the high mass ratio evolutionary track as shown along the top branch of Figure~\ref{fig:AsteroidFlowChart}.
Mutual body tides lead to synchronization of the spins to the orbit period and circularization of the orbit.
Tidal synchronization of each component of a high mass ratio binary occurs simultaneously, since they are of nearly equal size.
For ``rubble pile'' tidal parameters, these systems typically synchronize in  less than $10^5$ years \citep{Goldreich:2009ii,Jacobson:2011hp}, and so this process is considered an ``instantaneous'' process in the asteroid population evolution model.
This may be violated for high mass ratio systems, systems larger than $5$ km and with mass ratios $0.2 < q \lesssim 0.3$, which may take more than a million years to synchronize \citep{Jacobson:2011eq}.
Since high mass ratio systems have negative free energy, none of these systems can disrupt endogenously and all systems emerge as doubly synchronous binaries.
Once synchronous, the BYORP effect will expand or contract the mutual orbit.
Since this process can last for many millions of years, further evolution of high mass ratio binary systems is a long-term evolutionary process.

If the mass ratio of a system is determined to be low, then that system evolves along the low mass ratio evolutionary track as shown along the bottom branch of Figure~\ref{fig:AsteroidFlowChart}.
In \citet{Jacobson:2011eq}, this track is shown to immediately branch into four possible states, however all modeld chaotic ternary systems formed via secondary fission return to the chaotic binary state via escape of a member or impact between two of the members, so this track is not shown in Figure~\ref{fig:AsteroidFlowChart}.
Escape from low mass ratio systems is possible because they have positive free energy \citep{Scheeres:2009dc}, and \citet{Jacobson:2011eq} found numerically that $\sim 67\%$ of low mass ratio binaries do disrupt and form asteroid pairs as observed by \citet{Pravec:2010kt}.
Furthermore, \citet{Jacobson:2011eq} found that collisions between the two members occur in another $\sim 25\%$ of these systems forming re-shaped asteroids and that only $\sim8\%$ of low mass ratio binaries survive for more than $10^3$ years.

Typically, the secondary of these binaries synchronizes due to mutual body tidal dissipation in less than $10^5$ years \citep{Goldreich:2009ii,Jacobson:2011hp}, and so these binaries become singly synchronous systems within the ``instantaneous'' period of the asteroid population evolution model.
The model stochastically assigns an outcome to each rotationally fissioned low mass ratio system according to the probabilities reported above creating members of asteroid pairs, re-shaped asteroids, and singly synchronous binary systems.
Further evolution of singly synchronous binary systems due to the BYORP effect and tides is a long-term evolutionary process since the relevant timescales typically exceed a million years.

All resultant asteroid systems from both mass ratio regimes are propagated forward using the asteroid population evolution model with all of the asteroids that did not undergo rotational fission.
Members of asteroid pairs and re-shaped asteroids are subject to the YORP effect and collisions exactly as single asteroid systems that did not undergo rotational fission.
They are assigned new rotation rates from the original rotation rate distribution.
These systems are now single asteroid systems having complete one rotational fission lifetime cycle.
They can eventually rotationally fission again if they are accelerated to the appropriate rotational break-up speed of their size regime.

\subsection{``Long-term'' Binary Evolution}
\label{sec:longtermbinaryevolution}
Binary systems that have survived ``instantaneous'' evolution are treated differently than single systems in the asteroid population evolution model.
These systems are still subject to collisions as discussed in Section~\ref{sec:binariesandcollision}, and they would also be subject to YORP effect but not in the same way as single asteroids since the internal (i.e. spin and orbit states) evolution of binary systems is complicated by their mutual non-Keplerian gravity fields and mutual body tides.
Torques within binary systems such as the YORP effect and tides are generally much smaller than the BYORP effect \citep{Jacobson:2014hp} with the exception of those binaries that enter the tidal-BYORP equilibrium \citep{Jacobson:2011hp}, and so despite this complexity of multiple operating torques, an estimate of the lifetime of the binary can be estimated solely according to the BYORP effect evolution of the system.

The BYORP effect is an averaged torque on the orbit of synchronous satellites due to asymmetric emitted thermal radiation \citep{Cuk:2005hb,McMahon:2010by}.
The effect acts independently on each body, so that if both bodies are synchronous as in doubly synchronous binaries, then there is a BYORP torque on each, but for singly synchronous systems, the BYORP effect only acts on the synchronous secondary.
The direction of the BYORP torques is the fundamental parameter for determining the final evolutionary state of the system \citep{Cuk:2007gr,Jacobson:2011eq}.
The BYORP effect eventually destroys all doubly synchronous and half of all singly synchronous binary systems, as shown in Figure~\ref{fig:AsteroidFlowChart}.
The only exception to BYORP destruction are the singly synchronous systems which occupy an equilibrium between tides and the BYORP effect and are predicted to survive indefinitely unless there is exogenous interference such as a collision \citep{Jacobson:2011hp}.

\begin{\largetabletype}[!tb]
\centering
\begin{tabular}{|ccc|c|cccccc|c|}
\hline
\multicolumn{3}{|c|}{Evolutionary Track} & Likelihood & \multicolumn{7}{|c|}{Binary Lifetime Distributions} \\
$q$ & Direction & Aligned & Given $q$ & $\mu_\tau$ & $\mu_\tau$ & $\mu_\tau$ & $\mu_\tau$ & $\mu_\tau$ & $\mu_\tau$ & $\sigma_\tau$ \\
\hline
Low & Out & - & 0.5 & 4.88 & 5.88  & 6.88  & 7.88 & 8.88 & 9.88 & 0.71 \\
Low & In & - & 0.5 & $\infty$ & $\infty$  & $\infty$  & $\infty$  & $\infty$ & $\infty$ &  - \\
\hline
High & Out & No & 0.25 & 4.95 & 5.95  & 6.95  & 7.95  & 8.95  & 9.95 & 0.76  \\
High & Out & Yes & 0.25 & 4.61  & 5.61  & 6.61  & 7.61  & 8.61  & 9.61 & 0.55 \\
High & In & No & 0.25 & 4.42  & 5.42  & 6.42  & 7.42  & 8.42  & 9.42  & 0.75  \\
High & In & Yes & 0.25 & 4.09  & 5.09  & 6.09  & 7.09  & 8.09 & 9.09 & 0.55 \\
\hline
\hline
\multicolumn{4}{|r|}{BYORP Coefficient Distributions $\rightarrow$} & $\mu_B = -1$ &  $\mu_B = -2$ & $\mu_B = -3$ & $\mu_B = -4$ & $\mu_B = -5$ &$\mu_B = -6$ &  \\
\hline
\end{tabular}
\caption{Binary lifetime distributions for each binary evolutionary track and for each BYORP coefficient distribution.}
\label{tab:binarylifetimes}
\end{\largetabletype}

The asteroid population evolution model does not calculate the specific mutual orbit evolution of each binary system due to computational constraints.
Instead, each binary is assigned an evolutionary path determined by the system mass ratio and direction of the BYORP torque(s) in the system.
There are six distinct evolutionary paths as shown in Table~\ref{tab:binarylifetimes}: low mass ratio stable equilibrium with tides (contractive BYORP), low mass ratio expansive, high mass ratio expansive anti-aligned, high mass ratio expansive aligned, high mass ratio contractive anti-aligned, and high mass ratio contractive aligned.
Within each mass ratio regime, there is an equal likelihood to follow a specific track since there is nominally the same chance for a positive as negative BYORP coefficient and the BYORP coefficient of each body is independent of the other \citep{Cuk:2005hb,McMahon:2010jy}.
For instance, $25\%$ of high mass ratio systems evolve along the expansive track with aligned BYORP coefficients, since there is a $50\%$ chance that the primary will have a positive BYORP coefficient and a $50\%$ chance that the secondary will also have a positive BYORP coefficient.
Once the evolutionary track has been established for a binary system, it continues down that track for the rest of its lifetime.

The lifetime of a binary system is determined principally by the BYORP effect.
After synchronization of both members, tides may damp eccentricity from the system but do not strongly evolve the semi-major axis.
If only the secondary is synchronized, then tides are still important for contractive systems (i.e. the tidal-BYORP equilibrium) and while tides assist BYORP in expanding systems, tides are a strong function of semi-major axis and soon become much weaker than the BYORP effect.
There are also possible interruptions by exogenous processes (e.g. collisions, see Section~\ref{sec:binariesandcollision}).
The rate of expansion or contraction is determined primarily by the heliocentric orbit, absolute size of the system, and the BYORP coefficient. \citet{McMahon:2010jy} showed that to first order in eccentricity, the semi-major axis $a$ measured in primary radii $R_p$ evolves as:
\begin{equation}
\dot{a} = \frac{3 B_c}{2 \pi \omega_d \rho } \left( \frac{ a^{3/2} \sqrt{1+q}}{R_p^2 q} \right) \left( \frac{(2/3) F_\odot}{a_\odot^2 \sqrt{1 - e_\odot^2}} \right)
\end{equation}
where $B_c = B_p + B_s q^{2/3}$ is the combined BYORP coefficient.
The mass ratio $q^{2/3}$ factor is a direct result of the BYORP effect evolutionary equations \citep{McMahon:2010jy}.
For doubly synchronous systems, there is a BYORP coefficient for the primary $B_p$ and the secondary $B_s$, but for singly synchronous systems, there is only a BYORP torque on the secondary so the BYORP coefficient for the primary $B_p = 0$.
The BYORP coefficient is scaleless and depends solely on the shape of the synchronous member.

\subsubsection{BYORP coefficient distributions}
\begin{figure}[!tb]
\centering
\includegraphics[width=\columnwidth]{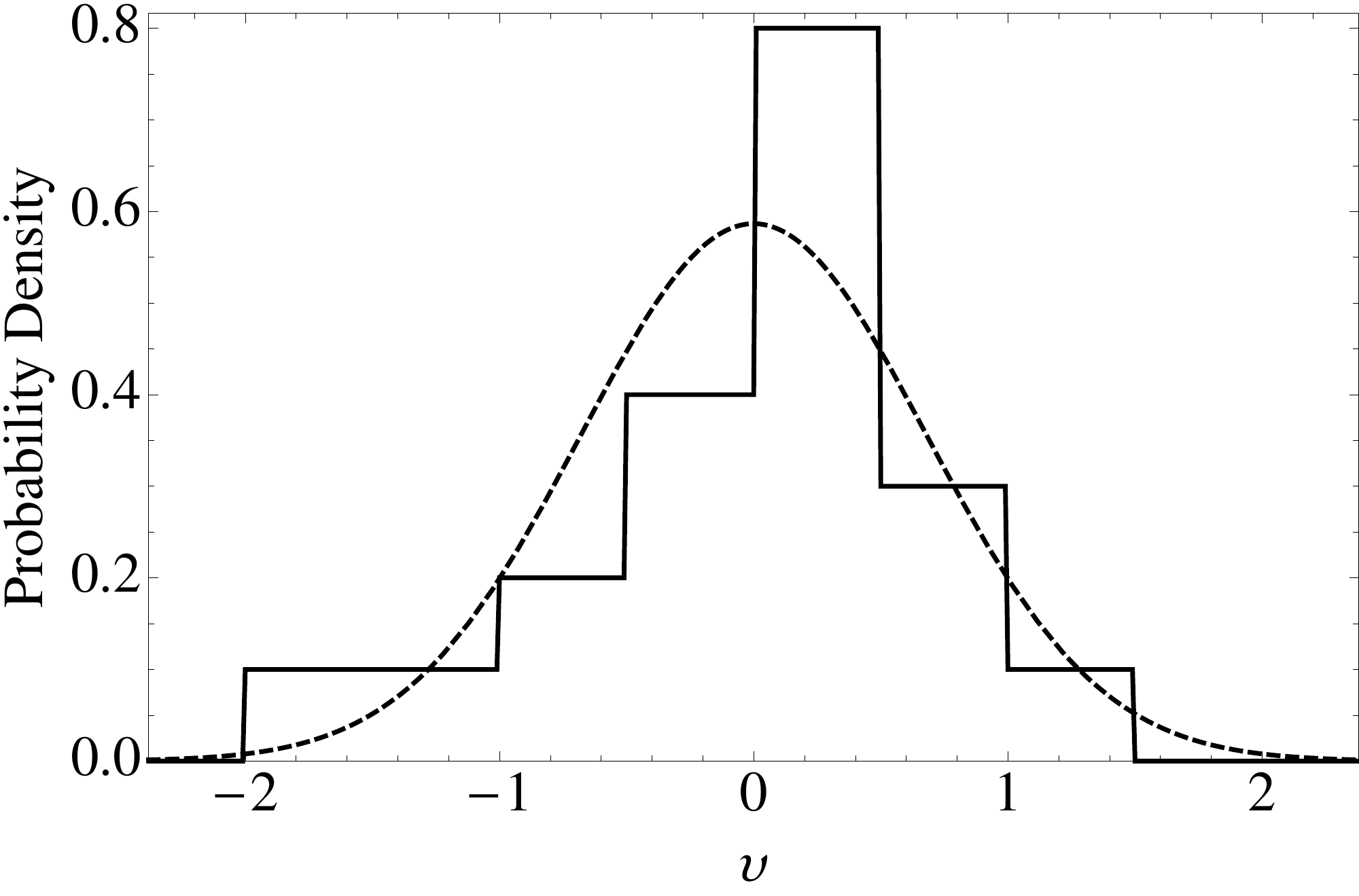}
\caption{A probability density histogram of $\upsilon$ of the observed singly synchronous population (bins are of width $0.5$).
The dashed line is the probability density function of a central normal distribution fit to the data where $\sigma_\upsilon = 0.68$.
Data is from \citet{Jacobson:2011hp}.}
\label{fig:BYORPDistribution}
\end{figure}
BYORP coefficients are determined solely by the shape of the asteroid, but determining the appropriate distribution of plausible BYORP coefficients is challenging.
The effect is similar to the detected Yarkovsky and YORP effects \citep{Chesley:2003bk,Taylor:2007kp,Lowry:2007by} and so the BYORP effect rests on strong theoretical support despite a lack of direct observation of BYORP-driven evolution.
The BYORP effect has never been directly measured, so a BYORP coefficient distribution cannot be derived from direct observation.
A detection may be precluded by the BYORP-tidal equilibrium hypothesis \cite{Jacobson:2011hp} and the possibly fast destruction of doubly synchronous binary systems \citep{Cuk:2007gr}.
Furthermore, there are very few well resolved asteroid shapes particularly of binary asteroid members.
The only current published BYORP prediction, \citet{McMahon:2010jy} estimated that $B_s = 2 \times 10^{-2}$ for the secondary of the 66391 (1999 KW$_4$) system using a vertice-and-facet shape model from \citet{Ostro:2006dq}.
This shape model is an order $8$ spherical harmonic representation with an average $26$ m facet edge length (corresponding to $7^\circ$ angular resolution).
Using this BYORP coefficient and the observed parameters of 66391, \citet{McMahon:2010jy} determined a Hill radius expansion timescale of $\sim5.4 \times 10^{4}$ years.
This expansion is very rapid compared to the typical YORP timescales of possible progenitors of $\sim10^6$ years assuming formation from YORP-induced rotational fission \citep{Rubincam:2000fg,Vokrouhlicky:2002cq,Capek:2004bl}.
Nominally, half of all synchronous binary asteroids are expected to expand due to the BYORP effect and 66391 may be a member of this population but observing this system as a binary rather than an asteroid pair is very unlikely given the difference between those two timescales.

This estimated BYORP coefficient also contradicts the BYORP-tidal equilibrium hypothesis in \citet{Jacobson:2011hp}, which states that the observable singly synchronous binary asteroids occupy an equilibrium between a contractive BYORP torque and the expansive mutual body tidal torque; this hypothesis requires a negative BYORP coefficient.
Further study by \citet[][pers. comm.]{McMahon:2012ty} concluded that the shape of 66391 should be known to a mean facet edge length of $8$ m (an angular resolution of $2.2^\circ$), using results scaled from an analysis of 25143 Itokawa (1998 SF$_{36}$), in order to model the BYORP coefficient with sufficient accuracy to prevent significant changes including sign changes.
For the related YORP effect, \citet{Statler:2009fw} concluded that spherical harmonic fits of order $\leq 10$ produce expected errors of order $100\%$ and for errors under $10\%$, the harmonic order of the fit must be at least $20$.
Furthermore, \citet{Statler:2009fw} showed that a crater half the object's radius can produce errors of several tens of percent; the observations of the secondary of 66391 did not uniformly cover the surface, a significant portion of the southern hemisphere is systematically not as accurate as the $7^\circ$ angular resolution of the rest of the model, and features such as craters may have not been observed \citep{Ostro:2006dq}.
Alarmingly, \citet{Rozitis:2012fq} conclude that the related YORP effect is very sensitive to surface roughness due to thermal-infrared beaming and that accurate YORP (and perhaps BYORP) coefficient estimation from shape models may require $1$ cm resolution.

\citet{Pravec:2010tc} determined that the direct detection of the BYORP effect and measurement of the BYORP coefficient would require multi-decade observations of small (semi-major axes of $<10$ primary radii and secondary radii $<1$ km) binaries.
Furthermore, this analysis did not include mutual body tides, which \citet{Jacobson:2011hp} predicted would create a stable equilibrium and halt mutual orbit evolution.
\citet{Scheirich:2015ez} conclude that for 175706 (1996 FG$_3$) this is true for at least this system.
Only the less numerous doubly synchronous systems do not have mutual body tides capable of creating the stable equilibrium.
69230 Hermes (1937 UB) is the smallest doubly synchronous system in both absolute size and heliocentric orbit, and is therefore the likeliest system for a direct detection of BYORP-driven orbit evolution.

While the hypothesized BYORP-tidal equilibrium prevents the direct measurement of the BYORP coefficients of singly synchronous binaries, it may be used to determine the relative distribution of BYORP coefficients.
In \citet{Jacobson:2011hp}, it is shown how for each system the balance between the BYORP and tidal torques determines the value of the combination of the  BYORP coefficient $B$ and the tidal parameters: tidal quality number divided by the tidal Love number $Q/k_p$ of the primary, degenerately:
\begin{equation}
\frac{BQ}{k_p} = \frac{2 \pi \omega_d^2 \rho R_p^2 q^{4/3} }{F_\odot a^7} a_\odot^2 \sqrt{1-e_\odot^2} = 2557 R_p\text{ km}^{-1}
\end{equation}
where the last equality is the fit to the singly synchronous binary data.

Since the BYORP coefficient is not a function of radius $R_p$, so that if the data is divided by a $Q/k_p \propto 2557 R_p$ km$^{-1}$ model, the resulting values reflect the distribution of BYORP coefficients $B$.
While this trick does not determine the absolute magnitude of the BYORP coefficient, it does provide information about the dispersion of the  BYORP coefficient distribution.
Figure~\ref{fig:BYORPDistribution} shows each system's normalized log BYORP coefficient $\upsilon$ as fit by a simple normal distribution with mean $\mu_\upsilon = 0$ and standard deviation $\sigma_\upsilon = 0.68$.
The observed distribution has a slight negative skew and a positive kurtosis compared to the normal distribution.
While the normalization of the singly synchronous data removed information about the absolute value of the BYORP coefficients, the standard deviation of those absolute coefficients is the same as the normalized coefficients so $\sigma_B = \sigma_\upsilon = 0.68$, where $\sigma_B$ is the standard deviation of $y$ and the absolute BYORP coefficients $B = 10^y$.

The mean $\mu_B$ of the distribution of $y$ is difficult to determine.
Estimating the absolute magnitude of the BYORP coefficient from  \citet{McMahon:2010jy} suggests a value for the mean of the distribution near $\mu_B = - 2$.
Even though this value is correct for the radar shape model of the secondary of 66391 rotated $180^\circ$ about either the radial or body axis orthogonal to the along track direction, however as discussed above, this estimation may not be accurate due to deficiencies of the shape model.

Since we cannot constrain the BYORP coefficient distribution, five different distributions are tested in the asteroid population evolution model: $\mu_B = -1$, $-2$, $-3$, $-4$, $-5$, and $-6$.
This is the second knob in the model; the other knob is the initial mass ratio fraction as described in Section~\ref{sec:massratiofraction}.
These BYORP coefficient distributions are used to generate the binary lifetime distributions that are then assigned to each binary system in the asteroid population evolution model.
Each BYORP distribution is tested independently and the entire asteroid population is then evolved from within the chosen distribution for the entirety of the run.

\subsubsection{Binary lifetime distributions}
\label{sec:binarylifetimedistributions}
The BYORP lifetime $\tau$ is determined by the evolution of the mutual orbit from a tidally synchronized semi-major axis to single member end states either re-shaped asteroids (e.g.
contact binaries) or asteroid pairs.
This evolution can be described as the evolution from an interior semi-major axis $a_\text{interior}$ to an exterior semi-major axis $a_\text{exterior}$ or vice versa:
\begin{align}
\tau =&  10^x R_p^2 a_\odot^2 \sqrt{1 - e_\odot^2} \\
x = & \log_{10}  \left[\frac{4 \pi \omega_d \rho q}{3 F_\odot B_c \sqrt{1 + q}} \left( \frac{1}{a_{interior}^{1/2}} - \frac{1}{a_{exterior}^{1/2}} \right) \right]
\label{eqn:lifetime}
\end{align}
where $F_\odot = 4.5 \times 10^{-5}$ g cm$^{-1}$ s$^{-2}$ is the solar constant at a $1$ AU circular orbit.
The BYORP lifetime $\tau$ is determined by the primary radius $R_p$, the heliocentric semi-major axis $a_\odot$ and eccentricity $e_\odot$, and $x$.
$x$ is the logarithm of all the other system parameter dependencies.
Rather than generating the necessary parameters to determine $x$ for each system, a million systems were generated outside of the asteroid population evolution model for each evolutionary path and the distribution of $x$ was determined.
Logarithmic normal distributions were fit to these generated distributions of $x$ with means of $\mu_\tau$ and standard deviations of $\sigma_\tau$.
Each distribution depends on the BYORP coefficients of the synchronous members, and the particular evolutionary track.
For each of the million systems, the BYORP coefficients are drawn from the distribution with the prescribed $\mu_B$ for that run.
Distributions of $x$ are shown in Figure 7, such that if $R_p$ is in km, $a_\odot$ is in AU, then $\tau$ is in years.

Each evolutionary pathway is defined by the sign of the BYORP coefficient for each synchronous member and the mass ratio of the system.
As mentioned earlier, the only evolutionary track that does not self-destruct is the BYORP contracting singly synchronous track.
These systems may contract or expand to some degree in semi-major axis, but the BYORP-tidal equilibrium hypothesis predicts that these systems reach a stable semi-major axis.
The interior $a_\text{interior}$ and exterior $a_\text{exterior}$ semi-major axes are given below for each of the evolutionary tracks in Table~\ref{tab:binarylifetimes}.

For high mass ratio doubly synchronous systems, the initial semi-major axis is always the tidally synchronized semi-major axis with the equivalent angular momentum as the rotational fissioned system at the time of fission.
Tidal dissipation will remove energy from the system, but angular momentum is conserved until the system is synchronized and the BYORP effect evolves the system.
This semi-major axis can be either the interior or exterior semi-major axis depending on the sign of the BYORP coefficient.
The initial semi-major axes for doubly synchronous systems $a_\text{d}$ is derived in~\ref{ref:derivationofinitialsemimajoraxesinbinaryevolution}.
This semi-major axis can be well approximated as a power law series expansion as a sole function of mass ratio $q$ and measured in primary radii $R_p$:
\begin{equation}
a_{doubly}  = 0.344 + \frac{0.00406}{q^3} + \frac{0.0132}{q^2} + \frac{0.815}{q} + 1.23 q
\end{equation}
For contracting high mass ratio systems, the interior semi-major axis $a_\text{interior}$ is contact between the two bodies:
\begin{equation}
a_\text{c} = 1+ q
\end{equation}
For both singly and doubly synchronous expanding systems, the exterior semi-major axis $a_{exterior}$ is the Hill radius $a_\text{Hill}$.
The Hill radius can be approximated in primary radii $R_p$:
\begin{equation}
a_{Hill} = q_\odot \left( \frac{4 \pi \rho}{9 M_\odot} \right)^{1/3}
\end{equation}
where $\rho = 2$ g cm$^{-3}$ is the density of the primary, $M_\odot  = 1.99 \times 10^{33}$ g is the mass of the Sun, and $q_\odot$ is the heliocentric perihelion of the barycenter of the system.
Asteroids at the outer edge of the Main Belt in circular orbits $q_\odot = 3.28$ AU have the largest Hill radii $a_\text{Hill} = 549$ primary radii and those at the inner edge in highly eccentric orbits with periapses just exterior to the Earth $q_\odot = 1$ AU have the smallest Hill radii $a_\text{Hill} = 168$ primary radii, but these radii are very large compared to the interior semi-major axes $a_\text{interior}$.
Since the BYORP lifetime is proportional to the difference between the inverse square roots of the interior and exterior semi-major axes, this factor of three difference in exterior semi-major axis translates to an at most $10\%$ difference in BYORP lifetime, if one extreme was chosen relative to the other.
To simplify the calculations, we use a single perihelion $q_\odot = 2.25$ AU, very close to the mean and median of the Main Belt Asteroid distribution.
This corresponds to a Hill radius $a_\text{Hill} = 377$ primary radii.

\subsection{Binaries and Collisions}
\label{sec:binariesandcollision}
If a binary participates in a catastrophic shattering collision then the binary is always destroyed.
This is determined by the same condition as a single asteroid from a comparison of the imparted specific kinetic energy and the critical impact specific energy.
Unlike single asteroids, cratering collisions can destroy a binary systems.
While these collisions by definition deliver less energy than the critical impact energy, these impacts can deliver enough energy to the system to disrupt the binary.
A simple condition for this disruption is a comparison of the delivered change in momentum to the system (delta V) and the escape velocity from the primary.
If the former exceeds the latter, then the system disrupts.

\subsection{Contact Binaries}
Contact binaries are formed from the merging of BYORP contracting high mass ratio binary systems.
These systems exist until either they undergo a rotational fission event or are subject to a catastrophic collision.
This is probably too optimistic a scenario since the binary system crosses a instability before contact \citep{Scheeres:2009dc}.
This instability causes the two components to begin to circulate and the orbit to evolve, but from simulations, these systems still collide and do so gently \citep{Jacobson:2011eq}.
These gentle collisions may be enough to reshape the new combined mass into a non-bifurcated shape that would not be easily identifiable as a contact binary.
The subjectivity of the contact binary label adds some uncertainty to the population statistics.

\section{Results of the asteroid population evolution model}
\label{sec:resultsoftheasteroidpopulationevolutionmodel}
The asteroid population evolution model produces a spin period distribution as a function of diameter similar to the observed population.
This is not of great surprise since the spin limit constraints were designed to reproduce the observed population and the model has been used successfully in the past for this purpose \citep{Marzari:2011dx}.
The model had two input parameters initial mass ratio fraction $F_i$ and mean of the log-normal distribution of BYORP coefficients $\mu_B$, and these inputs were permuted so that each combination produced a full set of model outputs.
We discuss each observable quantity output from the model and how that observable depends on the model free parameters: $F_i$ and $\mu_B$.
Combining all of the observables, we assemble a log-likelihood metric that can determine the best fit parameters.
Since the computational cost of running the asteroid population evolution model is high and we utilized a population of $2 \times 10^6$ asteroids, there is small variance when a particular set of input parameters is run a second time.
We use a Monte Carlo method to propagate the observed uncertainties to the comparison tests.
From the model, we identify a region where the free parameters can be well fit to the data.
They are discussed in detail in Section~\ref{sec:discussionoftheasteroidpopulationevolutionmodel}.

\subsection{Steady-State Binary Fraction}
\label{ref:steadystatebinaryfraction}
\begin{figure*}[!tb]
\centering
\includegraphics[width=8.9cm]{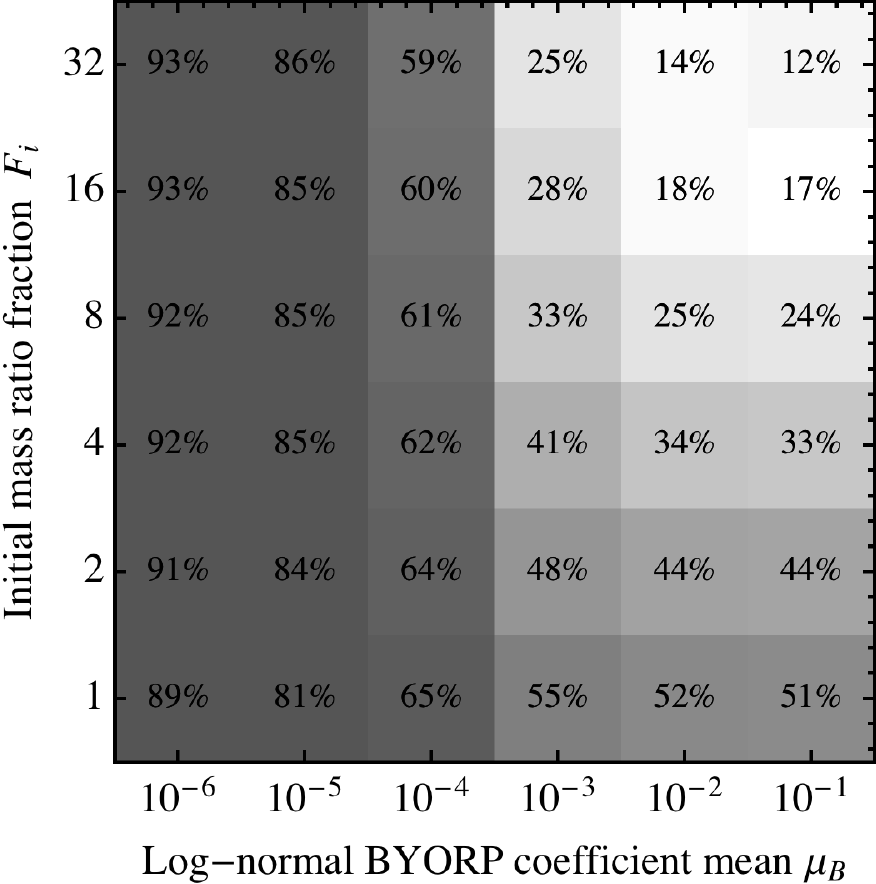}
\caption{The binary fraction $F_B$ of the model population is shown as a function of the two free parameters: the mean of the BYORP coefficient logarithmic normal distribution $\mu_B$ along the x-axis and the initial mass fraction $F_i$ along the y-axis.
Each grid point is determined from an independent run of the asteroid population evolution model with those values for the free parameters (otherwise the runs are identical).}
\label{fig:BinaryFractionPlot}
\end{figure*}
The asteroid population evolution model traces the evolution of a population with diameters from $200$ m to $20$ km.
However, observations typically do not go to such small sizes.
To replicate them, we will only consider asteroids with those diameters, which corresponds to an absolute magnitude $H \sim 21$ for typical asteroid albedos.

In Figure~\ref{fig:BinaryFractionPlot}, the steady-state binary fraction is shown as a function of both the initial mass ratio fraction and the log-normal BYORP coefficient distribution mean from asteroid population evolution model.
The difference between the asteroid population evolution model and the observations are shown as a heat map behind the model fractions (white indicates a close match).

Radar and photometric lightcurve observations supply independent and robust binary statistics regarding the near-Earth asteroid (NEA) population binary fraction, which we use as a proxy for the small Main Belt asteroid population (we discuss possible differences at the ends of the next paragraph).
Using radar observations, \citet{Margot:2002fe} reported that about $16\%$ of radar observed binary systems larger than $200$ m are binary systems.
Updated statistics from radar observations agree well with the better determined value of about $17\%$: $31$ binary systems out of $180$ asteroid systems with absolute magnitudes $H < 21$ approximate diameters $D \gtrsim 250$ m for an $p = 0.18$ albedo asteroid \citep{Taylor:2012vp}.
Photometric lightcurve analyses report a binary detection rate of $15 \pm 4\%$ for NEAs with diameters $D \gtrsim 300$ m and inferred mass ratios $q > 0.006$ \citep{Pravec:2006bc}.
This agrees with an initial assessment by \citep{Pravec:1999wt} that $17\%$ of near-Earth asteroid systems are binary.
The near-Earth asteroid population is significantly easier to observe than similar sized Main Belt asteroids, but for the sizes observed $D \lesssim 10$ km, rotational fission is expected to be the dominant formation mechanism.
For small diameter MBA systems $D \lesssim 10$ km, \citet{Pravec:2006vc} determine that there is a similar binary fraction in the inner Main Belt and this is supported by the results of the Binary Asteroid Photometric survey \citep{Pravec:2006bc,Pravec:2012fa}.
Tidal disruption of binary asteroids in the near-Earth asteroid population may lower the binary fraction in that population relative to the Main Belt \citep{Fang:2012go}.

Long binary lifetimes (small BYORP coefficients) naturally correspond to a high binary fraction.
A low initial mass ratio fraction has a higher binary fraction due to the more likely creation of synchronous long-lasting binary systems.
If we combine the photometric \citep{Pravec:2006bc} and radar \citep{Taylor:2012vp}  survey results and assume Poisson statistics for calculating the uncertainty, then the observed steady-state binary fraction is $16 \pm 6\%$.
The best parameter fits occur when the log-normal BYORP coefficient distribution mean is low, either $10^{-1}$ or $10^{-2}$ and the initial mass ratio fraction is greater than 8.

However, it is possible that comparing the main belt asteroid binary fraction to the near-Earth asteroid binary fraction may be misleading.
If an asteroid's average distance from the Sun during its orbit increases, then the rotational fission timescale increases as well since the YORP timescale would increase.
Since the rate of fission decreases, the creation of binary asteroids would slow controlling for all other factors other than average heliocentric distance.
When considering the steady-state population though, we must consider destruction as well as binary formation.
If the BYORP effect evolution is the dominant destructive route, then it scales identically with heliocentric distance as the YORP effect.
This is the primary reason, why it may be acceptable to use the near-Earth asteroid binary fraction as a proxy for the main belt asteroid binary fraction.
Furthermore, the YORP and BYORP timescales increase roughly by a factor of 10 from the near-Earth to main belt asteroid populations and this is the same factor by which the non-BYORP destructive routes increase---about 10 Myr for dynamical scattering into the Sun for the near-Earth asteroids and about 100 Myr for collisional disruption for the main belt asteroids.
From these considerations, we conclude that using the near-Earth asteroid binary fraction as a proxy is acceptable.

\subsection{Fast Binary Fraction}
\label{ref:fastbinaryfraction}

\begin{figure*}[!tb]
\centering
\includegraphics[width=8.9cm]{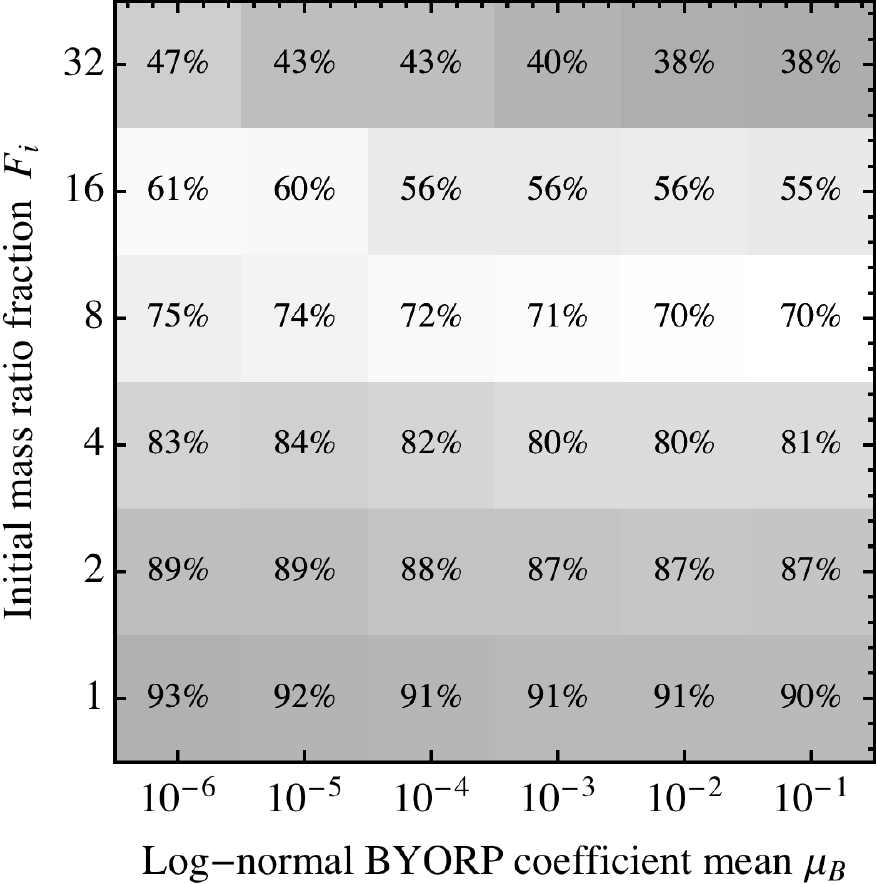}
\caption{The fast rotating fraction $F_F$ of the model population is shown as a function of the two free parameters: the mean of the BYORP coefficient logarithmic normal distribution $\mu_B$ along the x-axis and the initial mass fraction $F_i$ along the y-axis.
Each grid point is determined from an independent run of the asteroid population evolution model with those values for the free parameters (otherwise the runs are identical).}
\label{fig:LargeFastBinaryFractionPlot}
\end{figure*}

\citet{Pravec:2006bc} made a specific subpopulation observation that amongst fast-rotating binaries (spin periods between $2.2$ and $2.8$ hours) with diameters larger than $0.3$ km the binary fraction becomes $66 \pm 12\%$.
The asteroid population evolution model tracks the spin rate of single asteroids but since it does not evolve the system parameters of binaries, we rely on the binary evolution model to assume that all low mass ratio and no high mass ratio binaries will have rapidly rotating primaries \citep{Jacobson:2011eq}.

The fast rotating binary fraction as a function of the free parameters is shown in Figure~\ref{fig:LargeFastBinaryFractionPlot}.
Similar to the overall binary fraction, a large initial mass ratio fraction produces a small fast rotating binary fraction.
Unlike the overall binary fraction, the fast rotating binary fraction is not significantly dependent on binary lifetimes since only low mass ratio systems have rapidly rotating primaries.
There is a band around a initial mass fraction of $8$ that produces the smallest difference between the model and observation, however this constraint is softer than the overall binary fraction since the nearby bins have similar values.

\subsection{Steady-State Mass Ratio Fraction}
\label{ref:steadystatemassratiofraction}

\begin{figure*}[!tb]
\centering
\includegraphics[width=8.9cm]{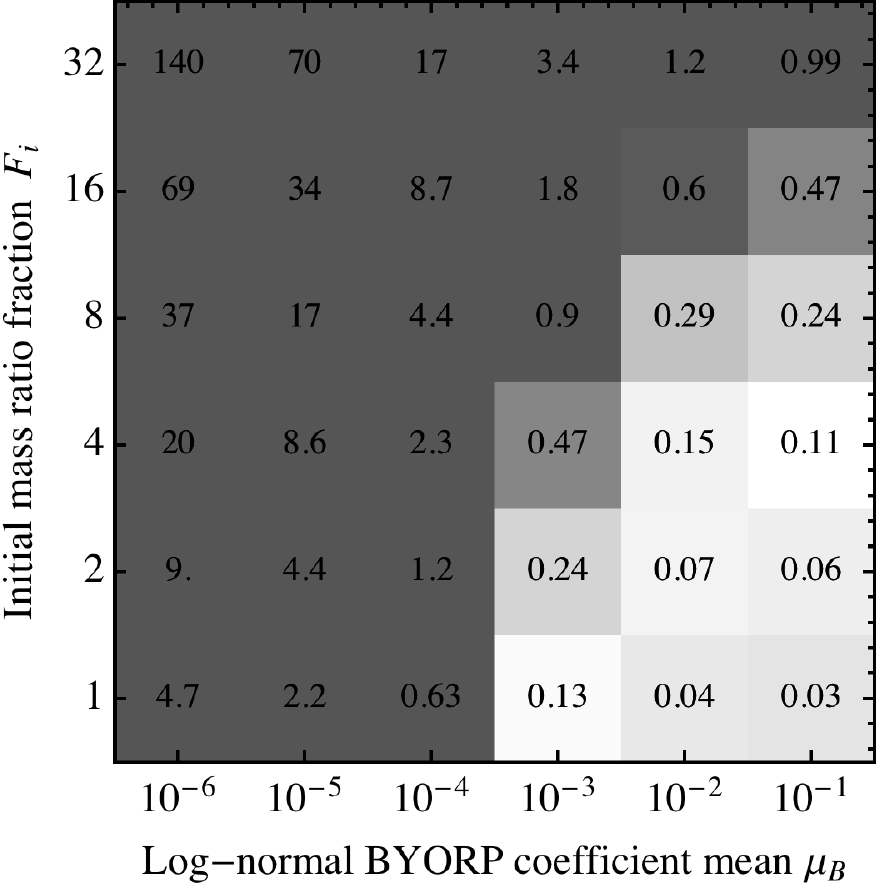}
\caption{The mass ratio fraction $F_q$ of the model population is shown as a function of the two free parameters: the mean of the BYORP coefficient logarithmic normal distribution $\mu_B$ along the x-axis and the initial mass fraction $F_i$ along the y-axis.
Each grid point is determined from an independent run of the asteroid population evolution model with those values for the free parameters (otherwise the runs are identical).}
\label{fig:MassRatioFractionPlot}
\end{figure*}

The steady-state mass ratio fraction is the evolved initial mass ratio fraction and the mass ratio fraction is the number of high mass ratio binaries divided by the number of low mass ratio binaries.
It is shown as a function of the free parameters in Figure~\ref{fig:MassRatioFractionPlot}.
Increasing the initial mass ratio fraction does increase the steady-state mass fraction, however that increase is mitigated when high mass ratio systems do not survive for as long as low mass ratio systems.
Also as the log-normal BYORP coefficient distribution mean decreases and binary lifetimes increase, the steady-state mass ratio fraction increases since the high mass ratio binaries are living longer relative to the low mass ratio synchronous systems, which are in a long-term equilibrium.

The binary asteroid catalogue provided by Pravec et al.
provides the best statistics regarding the steady-state mass ratio fraction.
This ratio is shown in Figure~\ref{fig:BinaryMassRatioHist} and is $0.11 \pm 0.08$ using Poisson statistics \citep{Pravec:2015uc}.
In Figure~\ref{fig:MassRatioFractionPlot}, the absolute difference between the model and the observation is shown as shading.
The best fits are a diagonal band from long binary lifetime and small initial mass ratio fractions to short binary lifetimes and high initial mass ratio fractions.
This is sensible trade-off in parameters to arrive at similar values for the steady-state mass ratio fraction.

\subsection{Contact Binary Fraction}

\begin{figure*}[!tb]
\centering
\includegraphics[width=8.9cm]{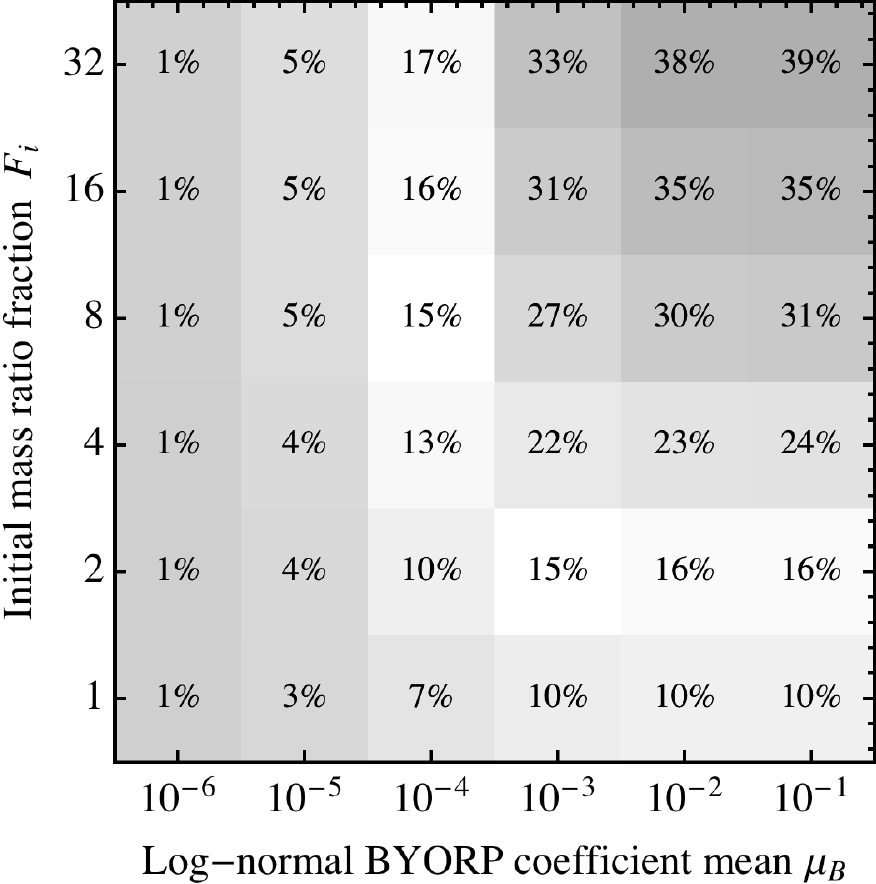}
\caption{The contact binary fraction $F_C$ of the model population is shown as a function of the two free parameters: the mean of the BYORP coefficient logarithmic normal distribution $\mu_B$ along the x-axis and the initial mass fraction $F_i$ along the y-axis.
Each grid point is determined from an independent run of the asteroid population evolution model with those values for the free parameters (otherwise the runs are identical).}
\label{fig:ContactFractionPlot}
\end{figure*}

In Figure~\ref{fig:ContactFractionPlot}, we show the model contact binary fraction as a function of the free parameters.
Contact binaries are formed from the destruction of inward evolving high mass ratio binaries, so when high mass ratio binaries are created often (large initial mass ratio fraction) and when they are destroyed frequently (large log-normal BYORP coefficient distribution mean), the contact binary fraction is high.

Only radar imaging can conclusively determine whether a system is a contact binary, but even then it is often a subjective result. \citet{Taylor:2012vp} provides the most recent estimate of $15 \pm 7\%$ using Poisson statistics.
This number is perhaps more likely to be an underestimate relative to the asteroid population evolution model definition of a contact binary because contact binary formation involves the low velocity collision of two asteroids and collision geometry and internal structure may dictate whether a collapsing high mass ratio system is observable as a contact binary.
In Figure~\ref{fig:ContactFractionPlot}, the absolute difference between the model and observations are shown.
If the model is over-counting contact binaries because the model always creates them at the end of the collapsing high mass ratio track evolution track, then the band of best fits would contract some about the upper right-hand corner and come more into agreement with the initial mass ratio fractions that the other observable constraints impose.

\subsection{Best Fit Parameters}
\begin{figure*}[!tb]
\centering
\includegraphics[width=8.9cm]{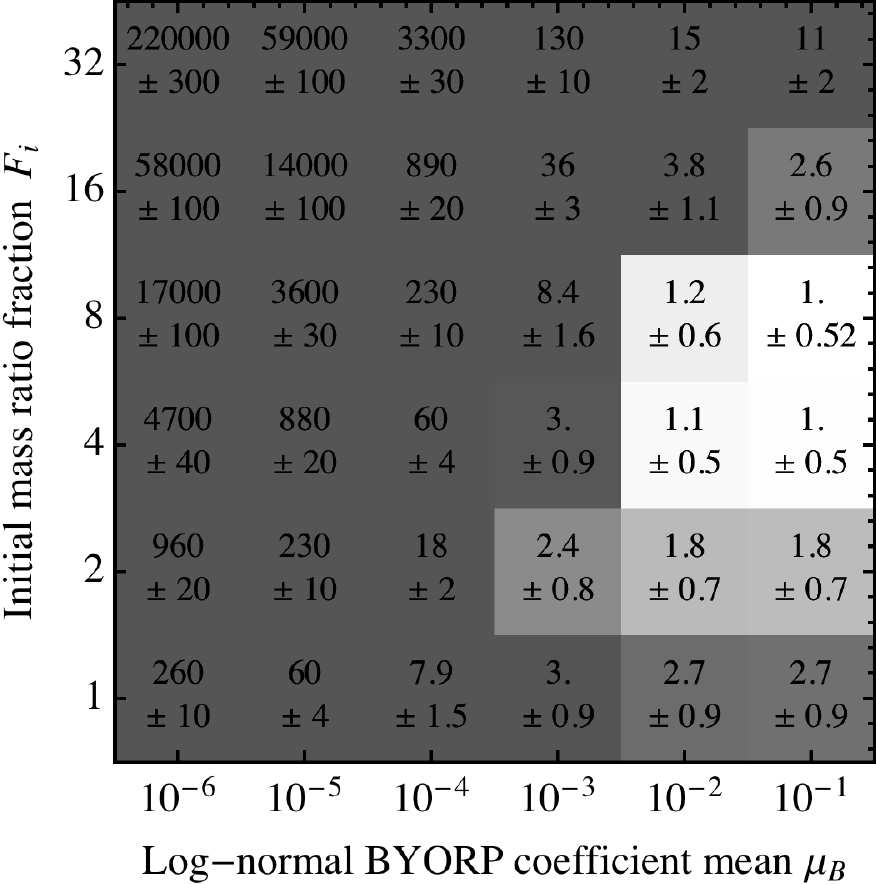}
\caption{The normalized log-likelihood is shown as a function of the free parameters of the asteroid population evolution model.
A low number is interpreted as more likely.}
\label{fig:LikelihoodThingPlot}
\end{figure*}

We can combine each of these observables into a single log-likelihood estimator for determining the best fit for the free parameters.
The log-likelihood metric we will use is a summation of the difference between the model output fraction $F_j$ for each observable $j$ and an observable fraction $F_{obs}$, which is drawn from a normal distribution with mean $\mu_j$ and standard deviation $\sigma_j$ in accordance with the values in the previous sections.

\begin{equation}
\mathcal{L} = A \sum_j \frac{ \left( F_j - F_{obs} \right)^2}{2 \sigma_j^2}
\end{equation}
A normalization is applied to make the best fit model have a value of 1.
The larger the normalized log-likelihood the less likely those set of parameters are.
Using Monte Carlo techniques, the uncertainty of the log-likelihood estimator can be determined.
It is important to note that due to computational constraints, the simulations are single runs and there is unaccounted for uncertainty.
Although, a few cases were run more than once and they were consistent with small changes to the reported values.
The log-likelihood metric is shown in Figure~\ref{fig:LikelihoodThingPlot}.

\subsection{Discussion of the asteroid population evolution model}
\label{sec:discussionoftheasteroidpopulationevolutionmodel}
The asteroid population evolution model identified a region in the phase space of the two free parameters in which the correct values are most likely to lie.
The log-normal BYORP coefficient distribution mean is likely to be greater than $-3$, which implies binary lifetimes less than $10^6$  years for systems that do not end up in the tidal-BYORP equilibrium.
This is similar to the formation and destruction cycle initially proposed by \citet{Cuk:2007gr} with the exception of the low mass ratio singly synchronous binaries, which we presume are captured in a tidal-BYORP equilibrium.
These short binary lifetimes are consistent with the understanding that the tight asynchronous population (e.g.
2004 $DC$) are newly formed binary systems that have yet to tidally relax.
However, \citet{Naidu:2015gp} demonstrate the possibility that simple tidal theory could dramatically underestimate the tidal locking timescale due to the spin-orbit coupling of a secondary's aspherical shape.
In this case, the model is incorrectly assuming that tidal synchronization can occur within the``instantaneous'' binary evolution time.

The best fit initial mass ratio fraction is $8$ but not statistically distinguishable from $4$. Because the mass ratio fraction is defined as the frequency of high mass ratio systems (0.2 to 1.0) over the frequency of low mass ratio systems (0.0 to 0.2), (note that the high mass ratio range is four times the extent of the low mass ratio range), the best fit initial mass ratio fraction is consistent with asteroids that fission nearly in half twice as frequently or at least as frequently as into two very unequal pieces.
The high mortality rate of low mass ratio systems in the ``instantaneous'' phase of binary formation is corrected by the synchronous low mass ratio binary population.
This is consistent with the hypothesis that asteroids are more likely to rotationally fission along interior planes and ``necks'' \citep{Sanchez:2012hz,Holsapple:2009tq} than from small events at the surface that accumulate in orbit into a larger satellite \citep{Walsh:2008gk}.

For these best fit parameters, the asteroid population evolution model provides some predictions regarding the Main Belt asteroid population.
The asteroid pair population is predicted to be about $2\%$ of the total population.
That is within the last $2$ Myrs, $2\%$ of the population was a member of a binary pair that disrupted.
These are mostly small asteroids, and it goes to less than $1\%$  for asteroids larger than a kilometer in diameter.

\section{Conclusions}
\label{sec:conclusions}
The YORP-induced rotational fission hypothesis predicts that the YORP effect rotationally accelerates asteroids until they fission and that this is the primary formation mechanism of binary asteroids.
We examine this hypothesis by modeling the main belt asteroid population between the sizes of 200 m and 20 km.
Our asteroid population evolution model rotationally evolves two million asteroids over 4.5 billion years according to the YORP effect and collisions.
Collisions can destroy both single and binary asteroids as well as modify the YORP coefficient after cratering collisions.
When these asteroids are rotationally accelerated to a rotational spin limit, they undergo rotational fission.
The outcome of each individual rotational fission event is drawn from statistical distributions as determined from the mutual orbit evolution model in \citet{Jacobson:2011eq}.
There are two important free parameters to the model, the initial binary mass ratio fraction $F_i$, which is the ratio of high to low mass ratio binaries created after a rotational fission event, and the strength of the BYORP effect $\mu_B$, which determines binary lifetimes.
Many binaries are ``instantaneously'' destroyed due to strong gravitational torques from spin-orbit coupling.
These form asteroid pairs, re-shaped asteroids if the mass ratio is low, and contact binaries if the mass ratio is high.
Those that survive evolve according to ``long-term'' effects such as the BYORP effect and tides.

The asteroid population evolution model utilizes a simplified form of the binary model described in \citet{Jacobson:2011eq}.
For instance, it ignores the formation of triple systems and wide asynchronous binaries.
Furthermore, the model utilizes evolutionary equations only accurate to first order in eccentricity.
The model asserts that singly synchronous binary systems are in a tidal-BYORP equilibrium.
This has the effect of scaling the strength of tidal evolution with the log-normal BYORP coefficient, which is chosen as a free parameter of the model.
If this assertion is incorrect, likely it is because the BYORP coefficient is significantly weaker than expected.
If it is significantly weaker, then a much lower initial binary mass fraction would be needed to explain the relative abundance of low mass fraction (typically singly synchronous) binaries.
However, this is unlikely to be the case since the theory behind the BYORP effect is robust especially given the observation of 1996 FG$_3$ within the tidal-BYORP equilibrium \citep{Scheirich:2015ez}.

The model also assumes an asteroid size distribution determined from collision evolution models which do not incorporate the YORP effect, however the YORP effect is expected to significantly deplete the population of small ($D \lesssim 10$ km) asteroids relative to the collisionally equilibrium size distribution \citep{Jacobson:2014bi}.
The inclusion of this effect may decrease the number of catastrophic and cratering events amongst the asteroid population.
This would allow more steady YORP effect evolution and probably lead to shorter periods between rotational fission events.
However, the YORP coefficient distribution used here already creates a spin rate distribution that matches that of the near-Earth and main belt asteroid populations \citep{Rossi:2009kz,Marzari:2011dx}.
Furthermore, the output of the asteroid population evolution model is compared to  observables, which are not absolute quantities but relative comparisons of sub-populations within the asteroid population, so the effect of a change of the absolute number of asteroids in a particular size bin may not be significant.

We compare the four outcomes from the model to observables: the steady-state binary fraction, the fast binary fraction, the binary mass ratio fraction and the contact binary fraction.
We find that the asteroid population evolution model can match each observable  individually and typically over a swath of parameter space.
When all of the observables are combined using a likelihood parameter, the model best fits all of the observables in only one location, so we determine that the best fit parameters are $F_i = 4$ or $8$ and $\mu_B = 10^{-1}$ or $10^{-2}$.
These best fit parameters are not very precise, but they are a unique global solution since each of the four observables carve out unique and generally orthogonal constraints on the parameter space.
Moreover, the best fit strengths of the BYORP effect match that predicted from a shape model.
Thus, we conclude that the YORP-induced rotational fission hypothesis can explain these four observables from a sophisticated asteroid population synthesis model.

\appendix
\section{Derivation of tidally synchronous semi-major axis for doubly synchronous binary evolution}
\label{ref:derivationofinitialsemimajoraxesinbinaryevolution}
Asteroids undergo rotational fission at some critical disruption rotation rate; this has been shown with analytic theory, observations of asteroid pairs, and computational numerics \citep{Scheeres:2007io, Pravec:2010kt, Sanchez:2012hz}.
This disruption rate and the shape of the asteroid at fission determine the angular momentum of the system during the ``instantaneous'' binary evolution stage identified in Section~\ref{sec:instantaneousbinaryevolution}.
In the doubly synchronous case, both bodies become synchronous with the orbit rate on similar short timescales but in the singly synchronous case, only the secondary is synchronized on a short timescale and the primary remains rotating at near the initial rate.
During this stage, energy is removed from the system via mutual body tidal dissipation but the angular momentum of the system is conserved.
By making some idealized approximations, the conservation of angular momentum is used to derive a tidal synchronization semi-major axis for the doubly synchronous systems $a_d$.

The angular momentum of an idealized binary system approximating each body as a constant density sphere is
\begin{equation}
H =  I_p \omega_{p} + I_s \omega_{s} + m a^2 \Omega
\end{equation}
where $I_n = 2 M_n R_n^2 / 5$ are the moments of inertia, $R_n$ are their radii, $M_n = 4 \pi \rho R_n^3 / 3$ are their masses, $m = M_p q / ( 1+q )$ is the reduced mass, $a$ is the distance between each body's center of mass, and $\Omega$ is the rotation rate about the system barycenter.
Additionally, the mass ratio is defined as $q = M_s / M_p = R_s^3 / R_p^3$ and the critical disruption rate for a specific mass ratio as $\omega_q = \sqrt{ (1+q)/(1+q^{1/3})^3}$.

In the idealized system described above, the initial angular momentum at the moment of rotational fission is a function of the mass ratio, the density and the primary radius.
Before entering into orbit, the two idealized components are initially separated only by their radii $a = R_p + R_s = R_p ( 1 + q^{1/3} )$.
All three rotation rates in the system are equivalent to the critical disruption rate for a specific mass ratio $\omega_{p} = \omega_{s} = \Omega = \omega_q $.
Therefore, the initial angular momentum of the system is
\begin{align}
H_i = & \frac{4 \pi \rho \omega_d R_p^5}{15} \sqrt{\frac{1+q}{\left(1 +q^{1/3}\right)^3}} \times \\
& \!\!\!\!\!\!\!\!\!\!\!\!\!\!\! \left( \frac{2 - 2 q^{1/3} + 2 q^{2/3} + 5 q + 5 q^{4/3} + 2 q^{5/3} - 2 q^2 + 2 q^{7/3}}{1 - q^{1/3} + q^{2/3}} \right) \nonumber
\end{align}
Doubly synchronous systems dissipate energy until all three rotation rates of the system are equivalent to the keplerian orbit rate $\omega_{p} = \omega_{s} = \Omega = \omega_{d} \sqrt{(1 +q )/ a_d^3 } $ where $a_d = a / R_p$ is the doubly synchronous synchronization semi-major axis normalized by the primary radius.
The synchronization angular momentum for a doubly synchronous is:
\begin{align}
H_d & = \frac{4 \pi \rho \omega_d R_p^5}{15} \times \\
& \left( \frac{ \left( 1 +q \right) \left( 2 + 2 \left( q + q^{5/3} + q^{8/3} \right) + 5 q a_d^2 \right)}{ \left( a_d \left( 1 + q^{1/3} \right) \left( 1 - q^{1/3} + q^{2/3} \right)  \right)^{3/2} } \right) \nonumber
\end{align}
Since angular momentum is conserved, $H_i = H_d$ and we obtain the synchronization semi-major axis $a_d$.
If we assume $a_d > 0$ and $ 0 \leq q \leq 1$, we can approximate the solution using a power series:
\begin{equation}
a_d  = 0.344 + \frac{0.00406}{q^3} + \frac{0.01322}{q^2} + \frac{0.815}{q} + 1.23 q
\end{equation}
is the initial tidally doubly synchronous semi-major axis measured in primary radii $R_p$.

\clearpage

\bibliographystyle{model2-names}
\bibliography{biblio}

\end{document}